\documentclass[12pt]{article}
\usepackage{graphicx,amsfonts,amsmath,amssymb,amsthm,mathrsfs,url,hyperref,tikz}
\usepackage{xcolor}

\title{Bohmian Trajectories for Hamiltonians with Interior--Boundary Conditions}

\author{
Detlef D\"urr\footnote{Mathematisches Institut, Ludwig-Maximilians-Universit\"at, 
     Theresienstr.~39, 80333~M\"unchen, Germany.
     E-mail: duerr@mathematik.uni-muenchen.de},\ 
Sheldon Goldstein\footnote{Departments of Mathematics, Physics and
     Philosophy, Rutgers University, Hill Center,  
     110 Frelinghuysen Road, Piscataway, NJ 08854-8019, USA.
     E-mail: oldstein@math.rutgers.edu},\
Stefan Teufel\footnote{Mathematisches Institut,
     Eberhard-Karls-Universit\"at, Auf der Morgenstelle 10, 72076
     T\"ubingen, Germany. E-mail:
     stefan.teufel@uni-tuebingen.de},\\
Roderich Tumulka\footnote{Mathematisches Institut,
     Eberhard-Karls-Universit\"at, Auf der Morgenstelle 10, 72076
     T\"ubingen, Germany. E-mail:
     roderich.tumulka@uni-tuebingen.de},\ and 
Nino Zangh\`\i\footnote{Dipartimento di Fisica dell'Universit\`a
     di Genova and INFN sezione di Genova, Via Dodecaneso 33, 16146 Genova, Italy.
     E-mail: zanghi@ge.infn.it}
}
\date{September 26, 2018}

\newcommand{\Hilbert}{\mathscr{H}}

\newcommand{\Q}{\mathcal{Q}}
\newcommand{\tQ}{\widetilde{\Q}}
\newcommand{\tmu}{\widetilde{\mu}}
\newcommand{\tH}{\widetilde{H}}
\newcommand{\tLaplace}{\widetilde{\Laplace}}

\renewcommand{\Im}{\mathrm{Im}}

\newcommand{\RRR}{\mathbb{R}}
\newcommand{\CCC}{\mathbb{C}}

\newcommand{\SSS}{\mathbb{S}}
\newcommand{\ZZZ}{\mathbb{Z}}

\newcommand{\scp}[2]{\langle #1|#2 \rangle}

\newcommand{\Laplace}{\Delta}

\newcommand{\orig}{{\mathrm{orig}}}

\newcommand{\va}{\boldsymbol{a}}
\newcommand{\vj}{\boldsymbol{j}}

\newcommand{\vq}{\boldsymbol{q}}

\newcommand{\vv}{\boldsymbol{v}}

\newcommand{\vx}{\boldsymbol{x}}
\newcommand{\vy}{\boldsymbol{y}}
\newcommand{\vz}{\boldsymbol{z}}
\newcommand{\vA}{\boldsymbol{A}}
\newcommand{\vX}{\boldsymbol{X}}
\newcommand{\vY}{\boldsymbol{Y}}

\newcommand{\valpha}{\boldsymbol{\alpha}}
\newcommand{\vomega}{\boldsymbol{\omega}}
\newcommand{\vzero}{\boldsymbol{0}}
\DeclareMathOperator{\Sym}{Sym}
\DeclareMathOperator{\Anti}{Anti}
\newcommand{\domain}{\mathscr{D}}

\newcommand{\be}{\begin{equation}}
\newcommand{\ee}{\end{equation}}
\newcounter{remarks}
\begin{document}
\maketitle
\begin{abstract}
Recently, there has been progress in developing interior-boundary conditions (IBCs) as a technique of avoiding the problem of ultraviolet divergence in non-relativistic quantum field theories while treating space as a continuum and electrons as point particles. An IBC can be expressed in the particle-position representation of a Fock vector $\psi$ as a condition on the values of $\psi$ on the set of collision configurations, and the corresponding Hamiltonian is defined on a domain of vectors satisfying this condition. We describe here how Bohmian mechanics can be extended to this type of Hamiltonian. In fact, part of the development of IBCs was inspired by the Bohmian picture. Particle creation and annihilation correspond to jumps in configuration space; the annihilation is deterministic and occurs when two particles (of the appropriate species) meet, whereas the creation is stochastic and occurs at a rate dictated by the demand for the equivariance of the $|\psi|^2$ distribution, time reversal symmetry, and the Markov property. The process is closely related to processes known as Bell-type quantum field theories.

\medskip

%  \noindent MSC (2000): 
%  81T10; %Model quantum field theories
%  58J32. %Boundary value problems on manifolds
\end{abstract}
\newpage
\tableofcontents

\section{Introduction}

A new type of Hamiltonian has recently been proposed \cite{TT15a} for quantum field theories (QFTs),
defined using \emph{interior--boundary conditions} (IBCs). See \cite{LP30, Mosh51a, Mosh51b, Mosh51c, ML91,Tho84,Yaf92,TG04} for earlier work involving IBCs, but with rather different purposes than in \cite{TT15a}. These Hamiltonians do not suffer from an ultraviolet (UV) divergence problem although they do not involve a UV cut-off such as would be provided by discretizing space or smearing out particles over a positive radius. On the contrary, in this new type of Hamiltonian, all particles are taken to have radius zero. The Hamiltonians involve particle creation and annihilation and have been shown \cite{ibc2a,LS18,Lam18}, for various examples of non-relativistic QFTs, to be free of a UV (or any other) divergence problem; viz., they have been shown to be rigorously defined and self-adjoint. 

An IBC is a type of boundary condition on the wave function that relates the value or derivative of the wave function on the boundary of configuration space to its value at a certain interior point. For particle creation, the relevant boundary consists of those configurations in which two particles meet at the same location. 

We define here the Bohmian trajectories naturally associated with such Hamiltonians and describe what they look like; that is, we develop an extension of Bohmian mechanics \cite{Bohm52,Gol01b,DT,Tum18} to Hamiltonians with IBCs. This extension is no longer deterministic but has the form of a stochastic Markov process in the appropriate configuration space, a process that is $|\psi_t|^2$-distributed at every time $t$.

\begin{figure}[h]
\begin{center}
\includegraphics[width=0.5\textwidth]{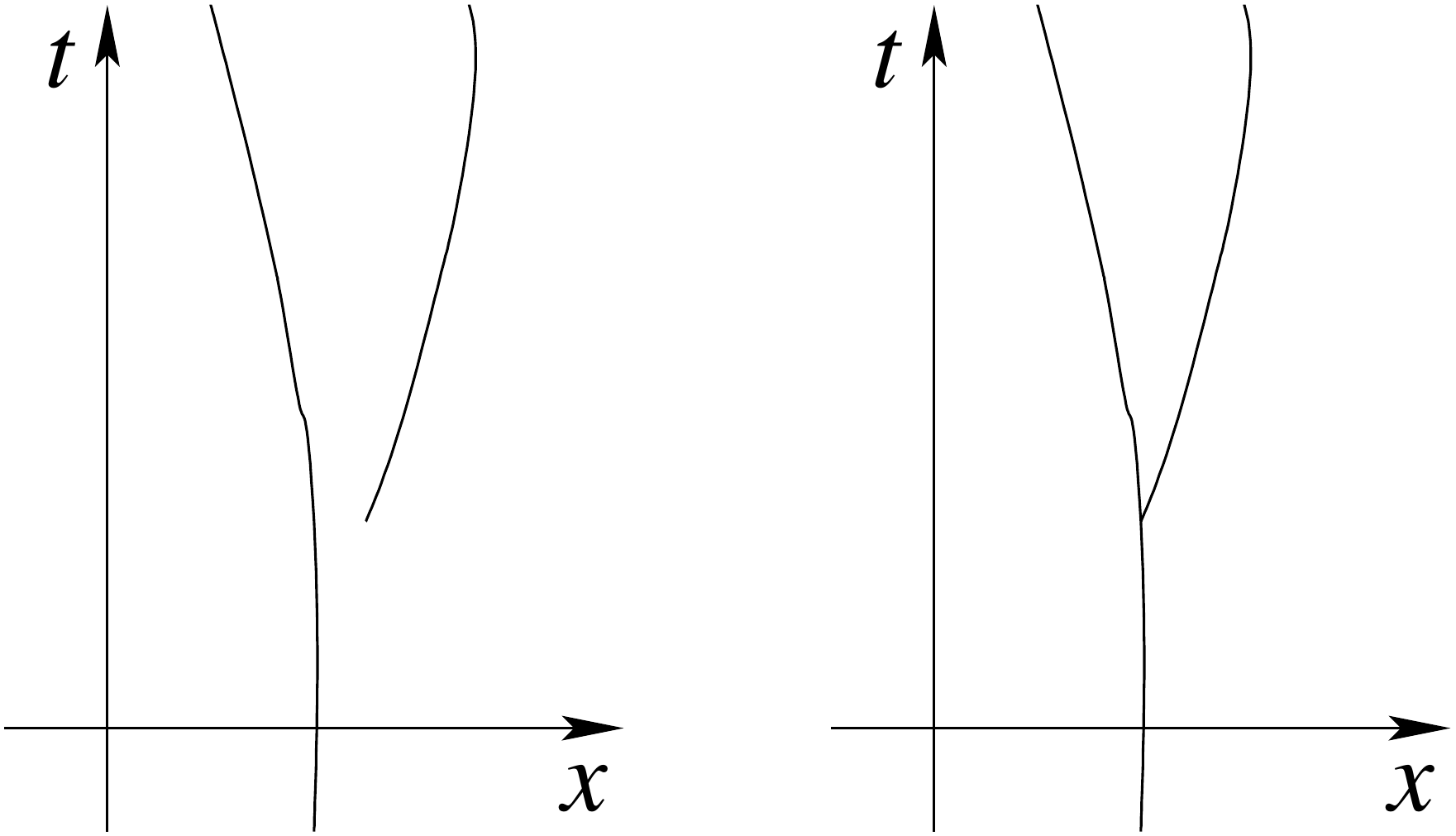}
\end{center}
\caption{Bohmian world lines in space-time for the emission of a $y$-particle from an $x$-particle in two kinds of models. LEFT: In a model with a UV cut-off, the $y$-particle gets created at a (small but) nonzero distance from the center of the $x$-particle. RIGHT: In a model with an interior--boundary condition, the $y$-particle gets created at the location of the $x$-particle, which has zero radius.}
\label{fig:radius}
\end{figure}

The theories that we develop here can be regarded as instances of \emph{Bell-type quantum field theories} \cite{Bell86,DGTZ03,DGTZ04,DGTZ05b}, versions of QFTs that, like Bohmian mechanics, provide particle trajectories; the possible paths in configuration space are piecewise solutions of Bohm's equation of motion, interrupted by jumps in configuration space, with the jump usually connected to the creation or annihilation of a particle. The QFTs considered in the references \cite{Bell86,DGTZ03,DGTZ04,DGTZ05b} just cited involved a UV cut-off, usually implemented by smearing out the particles over a positive radius $\delta$. For example, in a theory in which $x$-particles can emit and absorb $y$-particles, the emission of a $y$-particle corresponds to the occurence of a further ``Bohmian'' particle at any point within the $x$-particle, i.e., at any point within the 3-ball of radius $\delta$ around the center of the $x$-particle. In contrast, for Hamiltonians defined by means of an IBC, the $x$-particle has radius 0, and the $y$-particle gets created \emph{at the location} of the $x$-particle, see Figure~\ref{fig:radius}. Such a picture seems physically reasonable, in support of the IBC approach.

To study the combination of Bohm's trajectory picture with IBC Hamiltonians is useful from both perspectives: For the Bohmian picture, it provides an extension of the Bell-type QFTs known so far to a further class of Hamiltonians, indeed perhaps more reasonable and plausible Hamiltonians than the ones based on UV cut-offs. For the IBC approach, it provides a welcome visualization and clarification of the physical meaning of the Hamiltonians based on IBCs.

Preliminary considerations in the direction of this paper were reported in \cite{TG04,TT15b}. For an introduction to IBCs, see \cite{TT15b}; for a study of IBCs in 1 dimension, see \cite{KS16}; a brief overview of the results of this paper is given in Section 2 of \cite{EmQM17}; other recent and upcoming works on IBCs include \cite{Gal16,LN18,ST18,Sch18,IBCco1}. While the ideas of the IBC approach can also be applied to Dirac operators \cite{IBCdiracCo1}, we focus here on the non-relativistic case and give only a brief discussion of the analogous construction for the Dirac equation in Section~\ref{sec:Dirac}. It is a future goal to develop a model analogous to the ones described here for full quantum electrodynamics, with particle trajectories for electrons, positrons, and photons.
Other notable approaches to a version of QFTs with local beables (``hidden variables'') are based on either using configurations of infinitely many particles while avoiding the actual creation and annihilation of particles \cite{Col03a,Col03b,Col03c,CS07,DEO17} or using, instead of an actual particle configuration, an actual field configuration, see \cite{S10} and references therein, or assuming that fermions have beables but bosons do not \cite{Bell86} (or vice versa \cite{SW}); we will not consider these approaches here.

Like Bohmian mechanics and Bell-type QFTs, the models we describe here entail, as we will show on a non-rigorous level, that the actual configuration $Q_t$ at time $t$ is always $|\psi_t|^2$ distributed (and we then say that the process is \emph{equivariant}). Like Bell-type QFTs and unlike Bohmian mechanics (for a conserved number of particles), these models involve a stochastic motion of $Q_t$. We regard it as a serious possibility that the fundamental dynamical laws of physics may be stochastic in nature (i.e., that the time evolution may be inherently random). The main advantage of Bohmian mechanics is not so much its determinism as the clear picture of reality, independent of observation, that it provides.  

So, in the models developed in this paper, the path $t\mapsto Q_t$ is random, and thus a stochastic process (in fact, a Markov process), which we call the \emph{Bell-type process with IBC} because it naturally fits among the processes of Bell-type QFTs, or shorter the \emph{IBC process}. The stochastic element in the process is connected to the jumps; between jumps, the trajectory follows Bohm's deterministic equations of motion. For example, the emission of a $y$-particle by an $x$-particle (in the model of Section~\ref{sec:M2} below) occurs at a random time and in a random direction in space, with a probability distribution governed by the wave function according to one of the laws of the theory that we propose. 

This situation is similar to the one in Bell-type QFTs with a UV cut-off, where a $y$ particle is created at a random time and a random location within radius $\delta$ of the center of the $x$-particle. A difference is that the absorption event, which in Bell-type QFTs with UV cut-off is also stochastic (as it occurs at a random time), is deterministic in Bell-type processes with IBC: a $y$-particle gets annihilated when it hits an $x$-particle. While it may seem to break time reversal invariance that emission is stochastic and absorption deterministic, this is not so, as we will elucidate in Section~\ref{sec:reversal}. On the contrary, time reversal symmetry fixes uniquely the stochastic law governing the rate of particle creation. 
While Bell-type QFTs with UV cut-off can also obey time reversal symmetry, this symmetry does not dictate the rate for them, as several laws for the jump rate are compatible with it and with the quantum mechanical formula for the probability current, although one possibility for this law, the one chosen in Bell-type QFTs, is naturally selected by a minimality property. By the way, this choice of law now receives further support because it corresponds to the only possible law in IBC models. We also discuss how, in the limit of removing the UV cut-off (if the limit exists), the stochastic process $Q_t$ of a Bell-type QFT approaches the process introduced here for IBC models.

\begin{figure}[h]
\begin{center}
\includegraphics[width=0.4\textwidth]{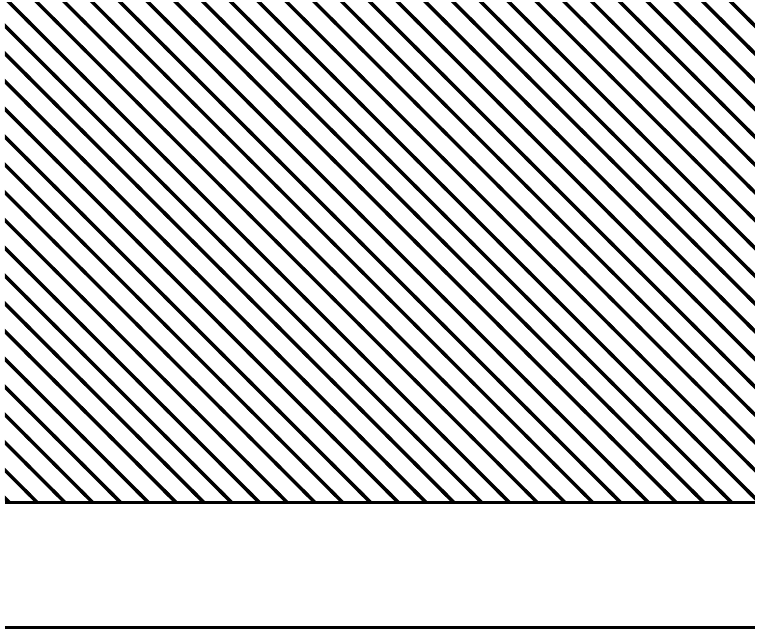}
\end{center}
\caption{The configuration space of Model~4 consists of a line and a half-plane.}
\label{fig:configM4}
\end{figure}

We will use four models for our discussion:

\begin{itemize}
\item \textit{Model~1.} This is a non-relativistic QFT involving two species of particles, $x$-particles and $y$-particles, both spinless and moving in $\RRR^3$, such that the $x$-particles can emit and absorb $y$-particles. The Hilbert space $\Hilbert$ is the tensor product of the fermionic Fock space of $L^2(\RRR^3,\CCC)$ for the $x$-particles and the bosonic Fock space of $L^2(\RRR^3,\CCC)$ for the $y$-particles. 

\item \textit{Model~2.} This is a simplified version of Model~1 in which the $x$-particles are fixed at certain locations in space, as would arise in the limit in which the mass of the $x$-particles tends to $\infty$. We consider here only the case of a single $x$-particle, and choose its location as the coordinate origin $\vzero\in\RRR^3$. So $y$-particles can be created and annihilated at the origin, and move around in between.

\item \textit{Model~3.} This model is a version of Model~2 that is further simplified by cutting off the sectors of the bosonic Fock space with 2 or more particles.

\item \textit{Model~4.} This model is even simpler and does not have much to do with particle creation any more. Its configuration space $\Q$ is the disjoint union of $\Q^{(1)}=\RRR$ and $\Q^{(2)}$ = the upper half-plane in $\RRR^2$ (see Figure~\ref{fig:configM4}); the boundary $\partial\Q$ of $\Q$, to which the IBC refers, is the horizontal axis in $\RRR^2$. Away from the boundary, the Hamiltonian is just the free Schr\"odinger operator.
\end{itemize}

We give a full definition of these models below, including the IBC approach to them. Model~1 and Model~2 were discussed, under these names, in \cite{TT15a}, and Models 3 and 4 were described in \cite{TT15b}. Model~1 is adapated from \cite{Lee54,Schw61,Nel64} (where similar models were considered without IBCs and without Bohmian trajectories), and also models like Model~2 have long been considered \cite{vH52,Der03} (without IBCs and without Bohmian trajectories), sometimes under the name ``Lee model.'' We will consider Models 1--4 in reverse order, the order of increasing complexity. For some of these models, the Hamiltonians of their IBC versions are known \cite{ibc2a,Lam18} to be bounded from below (as would be physically reasonable), although the IBC approach in general neither requires nor guarantees that Hamiltonians are bounded from below. 

The remainder of this paper is organized as follows. In Section~\ref{sec:M4}, we introduce and discuss the Bell-type process with IBC for Model~4. 
In Section~\ref{sec:M3M2}, we apply IBCs to particle creation and annihilation for Models 3, 2, and 1; we define the appropriate processes, show (non-rigorously) that they are equivariant, and compare the IBC approach to renormalization on the level of the Hamiltonians, the wave functions, and the Bell-type process. After Section~\ref{sec:M3M2}, we turn to more technical aspects of IBC processes. 
In Section~\ref{sec:sym}, we discuss the symmetries of the processes (particularly for Model~1), with particular attention to Galilean boosts. In Section~\ref{sec:Bell-type}, we compare the IBC process to the known processes (``Bell-type QFTs'' \cite{Bell86,Sudbery,Vink,DGTZ03,DGTZ04,DGTZ05b,Vi17}) for QFTs with UV cut-off. In Section~\ref{sec:lattice}, we formulate a lattice version of the IBC process and argue that in the continuum limit, the continuum version of the IBC process described in Section~\ref{sec:M4} is recovered. In Section~\ref{sec:co1}, we consider general IBC processes for codimension-1 boundaries (which are simpler than the physically realistic codimension-3 boundaries) and characterize such processes in general and abstract terms for Schr\"odinger and Dirac operators.

\section{Simple Example}
\label{sec:M4}

We begin with the simplest of our four models, Model~4. 

\subsection{Model~4 Comes First}

As mentioned, $\Q^{(1)}=\RRR$, $\Q^{(2)}=\{(x,y)\in\RRR^2: y\geq 0\}$, $\Q=\Q^{(1)}\cup \Q^{(2)}$, wave functions are complex-valued functions on $\Q$,  and volume in $\Q$ is understood as the measure $\mu$ defined by
\be\label{mudef}
\mu(S) = \mathrm{vol}_1(S\cap \Q^{(1)}) + \mathrm{vol}_2(S\cap \Q^{(2)})
\ee
for $S\subseteq \Q$, where $\mathrm{vol}_d$ means the $d$-dimensional volume (Lebesgue measure). For the restriction of a wave function $\psi$ to $\Q^{(n)}$ we write $\psi^{(n)}$; so, for a point $q\in\Q^{(n)}$, we can either write $\psi(q)$ or $\psi^{(n)}(q)$ (depending on whether we want to emphasize the number of the sector). The Hilbert space of the model is $\Hilbert=L^2(\Q,\CCC,\mu)$, whose inner product is
\be
\scp{\psi}{\phi} = \int\limits_{\Q^{(1)}} \!\!dx\, \psi^{(1)}(x)^*\, \phi^{(1)}(x) 
+ \int\limits_{\Q^{(2)}} \!\!dx\, dy\, \psi^{(2)}(x,y)^* \, \phi^{(2)}(x,y)\,.
\ee
The IBC reads \cite{TT15b}:
\be\label{IBC4}
\psi^{(2)}(x,0) = -\tfrac{2mg}{\hbar^2}\:\psi^{(1)}(x)
\ee
for every $x\in\RRR$. Here, $m>0$ is a mass parameter and $g\in\RRR$ a coupling constant.\footnote{The dimension of $g$ is (energy)$\times$(length)$^{3/2}$ if we take $\psi$ to have the dimension of the square root of a probability density, i.e., (length)$^{-1/2}$ for $\psi^{(1)}$ and (length)$^{-1}$ for $\psi^{(2)}$.} The corresponding Hamiltonian $H=H_{IBC}$ is:
\begin{subequations}\label{H4def}
\begin{align}
(H\psi)^{(1)}(x)\: &=-\tfrac{\hbar^2}{2m} \partial^2_x \psi^{(1)}(x) + g\, \partial_y \psi^{(2)}(x,0) \label{H4def1} \\
(H\psi)^{(2)}(x,y)\: & =-\tfrac{\hbar^2}{2m} \Bigl(\partial^2_x+\partial_y^2\Bigr) \psi^{(2)}(x,y) \quad \text{for }y>0\,.\label{H4def2}
\end{align} 
\end{subequations}
(The reasons for setting up the equations this way will become clearer once we have described the Bell-type process and its probability current.)
One can show that $H$ is self-adjoint on a suitable dense domain in $\Hilbert$ consisting of functions satisfying the IBC \eqref{IBC4}, so that $e^{-iHt/\hbar}$ is a unitary operator on $\Hilbert$, and $\psi_t=e^{-iHt/\hbar}\psi_0$ is the solution of the Schr\"odinger equation
\be\label{Schr1}
i\hbar \frac{\partial \psi_t}{\partial t} = H\psi_t\,.
\ee
We will assume in the following that $\psi_0$ (and thus also $\psi_t$) lies in the domain of $H$ and in particular satisfies the IBC.

\subsection{Process for Model~4}
\label{sec:Q4def}

The Bell-type process for this model, for any solution $\psi_t$ to \eqref{Schr1}, is defined as follows. The initial configuration $Q_0$ is chosen with the $|\psi_0|^2$ distribution. At any time $t$, if $Q_t$ lies in the interior of $\Q^{(2)}$, then it moves according to Bohm's equation of motion \cite{Bohm52,Gol01b,DT}, 
\be\label{Bohm4}
\frac{dQ_t}{dt} = \tfrac{\hbar}{m} \, \Im \, \frac{\nabla \psi_t}{\psi_t}(Q_t)\,,
\ee
or, equivalently,
\be\label{Bohm4b}
\frac{dQ_t}{dt} = \frac{j}{\rho}(Q_t)
\ee
in terms of the quantities
\begin{subequations}
\begin{align}
j&= \tfrac{\hbar}{m} \, \Im (\psi^*\nabla \psi) \label{jdef}\\
\rho &= |\psi|^2\label{rhodef}
\end{align}
\end{subequations}
that are usually called the probability current and probability density in quantum mechanics (and that will turn out to indeed be the probability current and density for our process).

As soon as the configuration hits the boundary $\partial \Q = \partial \Q^{(2)}=\{y=0\}$, say at $(x,0)$, it jumps to $x\in\Q^{(1)}$, and continues moving there according to Bohm's equation of motion \eqref{Bohm4}, now understood on $\Q^{(1)}$. The motion in $\Q^{(1)}$ will be interrupted at a random time $T$ whose distribution is specified below. At time $T$ the configuration jumps from $\Q^{(1)}$ to the boundary of $\Q^{(2)}$; if $X$ is the position immediately before the jump,\footnote{The notation $t\nearrow T$ means the limit $t\to T$ with $t<T$; $t\searrow T$ means $t\to T$ with $t>T$.}  $X=\lim_{t \nearrow T} Q_t$, then the point it jumps to is $\lim_{t\searrow T} Q_t=(X,0)$. After the jump, the configuration moves again according to Bohm's equation of motion \eqref{Bohm4}, etc. The distribution of $T$ can be expressed by specifying the jump rate $\sigma$, i.e., the probability per time of a jump to occur: the probability of a jump between $t$ and $t+dt$, given that $Q_t=x\in \Q^{(1)}$, is
\be
\sigma_t \bigl(x\to(x,0)\bigr) \, dt\,.
\ee
It is one of the laws of the theory that the jump rate is
\be\label{jumprate4}
\sigma_t\bigl(x\to(x,0)\bigr) = \tfrac{\hbar}{m} \, \frac{\Im^+ \bigl[ \psi^{(2)}(x,0)^* \; \partial_y \psi^{(2)}(x,0) \bigr]}{|\psi^{(1)}(x)|^2}
\ee
with the notation
\be\label{+def}
s^+=\max\{0,s\}
\ee
for the positive part of $s\in\RRR$.
(We write $\psi^{(1)}(x)$ for $\psi(x)$ to emphasize that we are evaluating $\psi$ in sector $\Q^{(1)}$.) 

\begin{figure}[h]
\begin{center}
\includegraphics[width=0.4\textwidth]{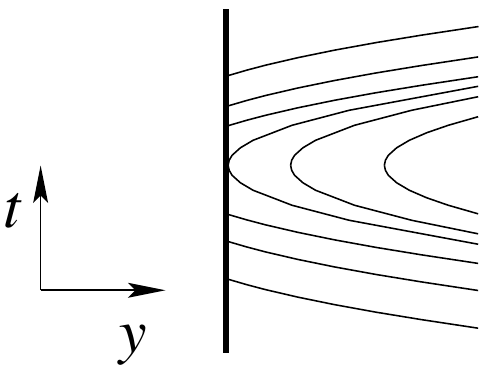}
\end{center}
\caption{Several Bohmian trajectories (i.e., solutions of \eqref{Bohm4} in $\Q^{(2)}$), some of which begin or end at the boundary point $(x,0)$; a $yt$-diagram is shown, depicting only the $y$-component of the trajectories as a function of $t$ (upward); the vertical bar represents the boundary $y=0$. The uppermost three trajectories begin on the boundary and have $j_y^{(2)}(x,0,t)>0$ at the initial point; the lowermost three end on the boundary and have $j_y^{(2)}(x,0,t)<0$ at the final point.}
\label{fig:beginning}
\end{figure}

For the law \eqref{jumprate4} to be meaningful and sufficient, we need that whenever the process jumps to $(x,0)$, a unique trajectory begins there and leads away from the boundary. This is in fact the case: Since the jump rate \eqref{jumprate4} can be written as 
\be\label{jumprate4b}
\sigma_t\bigl(x\to(x,0)\bigr) = \frac{j^{(2)}_y(x,0)^+}{\rho^{(1)}(x)}\,,
\ee
where $j_y$ means the $y$-component of the current in $\Q^{(2)}$, the process can only jump to $(x,0)$ if the current $j^{(2)}(x,0)$ has positive $y$-component, and thus points away from the boundary. But then also the Bohmian velocity \eqref{Bohm4b} points away from the boundary (i.e., has positive $y$-component), and there is a unique solution of the ODE \eqref{Bohm4b} beginning at $(x,0)$; see Figure~\ref{fig:beginning}. In contrast, if $j_y^{(2)}(x,0)<0$, then a solution moving towards the boundary ends there, so it would not be possible to jump to $(x,0)$ and then move along the trajectory passing through $(x,0)$ at that time. The case $j_y^{(2)}(x,0)=0$ is more complicated but irrelevant here because the jump rate \eqref{jumprate4b} vanishes in this case. What if $\psi^{(1)}(x)=0$? Then the jump rate \eqref{jumprate4} is ill defined, but this is not a problem because the process should be expected to have probability zero to ever reach such an $x$ \cite{bmex,GT05,TG04,TT05}.

This completes the definition of the process $(Q_t)_{t\geq 0}$ (or just $Q_t$ for short). It is clear that $Q_t$ is a \emph{Markov process}, i.e., a stochastic process for which the probability distribution of the future path depends on the past only through the present configuration. (Bohmian mechanics, the known Bell-type QFTs \cite{DGTZ05b}, and the analogous process on a graph \cite{Tum05} are also Markov processes.)

\subsection{Equivariance}
\label{sec:equiv}

We now turn to the derivation of the equivariance of the $|\psi|^2$ distribution (i.e., of the process defined in Section~\ref{sec:Q4def}). 

Consider an arbitrary probability density (instead of $|\psi|^2$) for the initial configuration $Q_0$, and let this density be denoted by $\rho$. Then we can formulate the transport equations for $\rho$, or probability balance, as follows. In the interior of $\Q^{(2)}$, $\rho$ gets transported according to the continuity equation
\be\label{continuity2}
\frac{\partial \rho^{(2)}}{\partial t}(x,y) 
= -\partial_x \Bigl(\rho^{(2)} \, v^{(2)}_x\Bigr)
- \partial_y \Bigl(\rho^{(2)}\, v^{(2)}_y\Bigr)\,,
\ee
where $v^{(2)}=(v^{(2)}_x,v^{(2)}_y)$ is the Bohmian velocity vector field,
\be\label{v2def}
v^{(2)}=j^{(2)}/|\psi^{(2)}|^2\,.
\ee
Some amount of $\rho^{(2)}$ gets lost due to trajectories that hit the boundary and jump to $\Q^{(1)}$. In $\Q^{(1)}$, $\rho$ evolves according to
\be\label{continuity1}
\frac{\partial \rho^{(1)}}{\partial t}(x) 
= -\partial_x \Bigl(\rho^{(1)}\, v^{(1)}\Bigr) 
+ \rho^{(2)}(x,0)\,\bigl[- v^{(2)}_y(x,0)\bigr]^+ 
- \rho^{(1)}(x)\, \sigma_t\bigl(x\to(x,0)\bigr) \,.
\ee
The first term on the right-hand side represents the change in $\rho^{(1)}(x)$ due to transport of $\rho^{(1)}$ along $\Q^{(1)}$, the second term represents the gain due to jumps coming from $(x,0)$, and the third the loss due to jumps from $x$ to $(x,0)$. Note that $\rho^{(2)}(x,0)\, \bigl[ -v^{(2)}_y(x,0)\bigr]^+ \, dx\, dt$ is the amount of probability arriving at $\partial \Q^{(2)}$ due to motion in $\Q^{(2)}$ between $(x,0)$ and $(x+dx,0)$.

Now the following equations for the time evolution of $|\psi|^2$ follow from the Schr\"odinger equation with the Hamiltonian \eqref{H4def}. In the interior of $\Q^{(2)}$,
\be\label{psi22}
\frac{\partial |\psi^{(2)}|^2}{\partial t}(x,y) = - \partial_x j_x^{(2)} - \partial_y j_y^{(2)}
\ee 
with $j$ as in \eqref{jdef}, and in $\Q^{(1)}$,
\be\label{psi12}
\frac{\partial |\psi^{(1)}|^2}{\partial t}(x) = -\partial_x j^{(1)} + \tfrac{2}{\hbar}\, \Im \bigl[ \psi^{(1)}(x)^*\, g\, \partial_y \psi^{(2)}(x,0)  \bigr]\,.
\ee
The IBC \eqref{IBC4} allows us to replace $\psi^{(1)}(x)^*$ on the right-hand side by $-(\hbar^2/2mg) \, \psi^{(2)}(x,0)^*$, so we obtain that
\be\label{psi12ibc}
\frac{\partial |\psi^{(1)}|^2}{\partial t}(x) = -\partial_x j^{(1)} - j_y^{(2)}(x,0)\,.
\ee
Thus, it follows from \eqref{jumprate4} and \eqref{v2def} that whenever $\rho=|\psi|^2$, the right-hand side of \eqref{continuity2} agrees with that of \eqref{psi22}, and the right-hand side of \eqref{continuity1} agrees with that of \eqref{psi12ibc}. 
Thus, $\rho=|\psi|^2$ is a solution to \eqref{continuity2} and \eqref{continuity1}, establishing the equivariance of the $|\psi|^2$ distribution.

This calculation also conveys how the conservation of $|\psi|^2$ works for this Hamiltonian: The second term on the right-hand side of \eqref{H4def1} ensures, together with the IBC \eqref{IBC4}, that the continuity equation \eqref{psi12ibc} for $|\psi^{(1)}|^2$ contains an additional term (the second term on the right-hand side) that compensates exactly the loss of $|\psi^{(2)}|^2$ due to flux into the boundary while yielding the gain of $|\psi^{(2)}|^2$ due to the jumps described by \eqref{jumprate4}.

\subsection{Remarks}
\label{sec:M4rem}

\begin{enumerate}
\setcounter{enumi}{\theremarks}

\item\label{rem:QT} \textit{At the time of jump.} 
Two types of jumps occur: $(x,0)\to x$ (deterministic) or $x\to (x,0)$ (stochastic). Let $T$ denote the time of either jump.  We have not specified whether $Q_T=x$ or $Q_T=(x,0)$. For the sake of a complete mathematical definition of the process, various choices for $Q_T$ can be adopted. For example, we could define that always $Q_T=x$ (in the lower sector), or that always $Q_T=(x,0)$ (in the upper sector). Both choices define Markov processes, and the differences between them seem physically irrelevant.

\item\label{rem:Qcirc} \textit{Choice concerning the boundary.} Another mathematical fine point that is physically irrelevant is whether boundary points should be regarded as elements of $\Q$ or not. For clarity, let us write $\Q^\circ=\Q\setminus \partial \Q$ for the \emph{interior} of $\Q$ (the set of non-boundary points) and $\overline{\Q}=\Q\cup \partial \Q$ for the \emph{completion} of $\Q$ (the set of boundary \emph{and} non-boundary points).\footnote{The definition of $\Q^\circ$ should not be conflated with the definition of the interior of a subset in a topological space, as the interior of the whole space is always the whole space.} Above, we took $\Q=\overline{\Q}$, but we could equally well have defined $\Q^{(2)}=\{(x,y)\in\RRR^2:y>0\}$ (with a $>$ sign instead of $\geq$), which would have led to $\Q=\Q^\circ$. Since $\partial\Q$ is a $\mu$-null set, we have that $L^2(\Q^\circ,\CCC,\mu)=L^2(\overline{\Q},\CCC,\mu)$. Since $\psi$ is then not defined on $\partial\Q$, we would have to write $\lim_{y\searrow 0}\psi^{(2)}(x,y)$ instead of $\psi^{(2)}(x,0)$ in \eqref{IBC4} and \eqref{jumprate4} and $\lim_{y\searrow 0} \partial_y \psi^{(2)}(x,y)$ instead of $\partial_y \psi^{(2)}(x,0)$ in \eqref{H4def1} and \eqref{jumprate4}. Moreover, since we want that $Q_t\in\Q$ for all $t$, we would need to demand that $Q_T$ lies in the lower sector (see Remark~\ref{rem:QT}). Except possibly for the choice of $Q_T$, the IBC process is the same as for the previous choice $\Q=\overline{\Q}$, as every trajectory in $\Q^{(2)}$ that hits the boundary has a unique limiting arrival point $\lim_{t\nearrow T} Q_t\in\partial \Q$, and conversely, there is no more than one trajectory whose limit backwards in time at a given time $T$, $\lim_{t\searrow T}Q_t$, is a given boundary point. Thus, the choice $\Q=\Q^\circ$ vs.\ $\Q=\overline{\Q}$ actually does not matter.

Moreover, if we wish, we can even take the wave function $\psi$ to be defined on $Q^\circ$ and the process $(Q_t)$ to move in $\overline{\Q}$. This choice will be convenient in Section~\ref{sec:M3M2}.

\item\label{rem:radical} \textit{Another topology on configuration space.} One can take a somewhat different view of the same process by introducing a different topology on $\Q^\circ$, which we call the \emph{radical topology}. It is obtained by identifying $\Q^{(1)}$ with $\partial \Q^{(2)}$, viz., $x$ with $(x,0)$. This means, for example, that an open neighborhood of $x$ contains not only nearby points in $\Q^{(1)}$ but also points near $(x,0)$ in $\Q^{(2)}$.\footnote{So in a sense, the topology is not very radical at all: In the present example (Model 4), it is the standard topology of a closed half plane. The name ``radical topology'' should not be over-interpreted.}
In this topology, $\Q^\circ$ is a connected space, and the process $Q_t$ has continuous paths. Since $\Q^{(1)}$ can now be pictured as the $y=0$ line in $\RRR^2$, a typical path starting with $y>0$ may reach $y=0$ sooner or later, stay on the $y=0$ line for a random duration, then leave that line into the upper half plane, etc.; see Figure~\ref{fig:radical}. 

\begin{figure}[h]
\begin{center}
\includegraphics[width=0.5\textwidth]{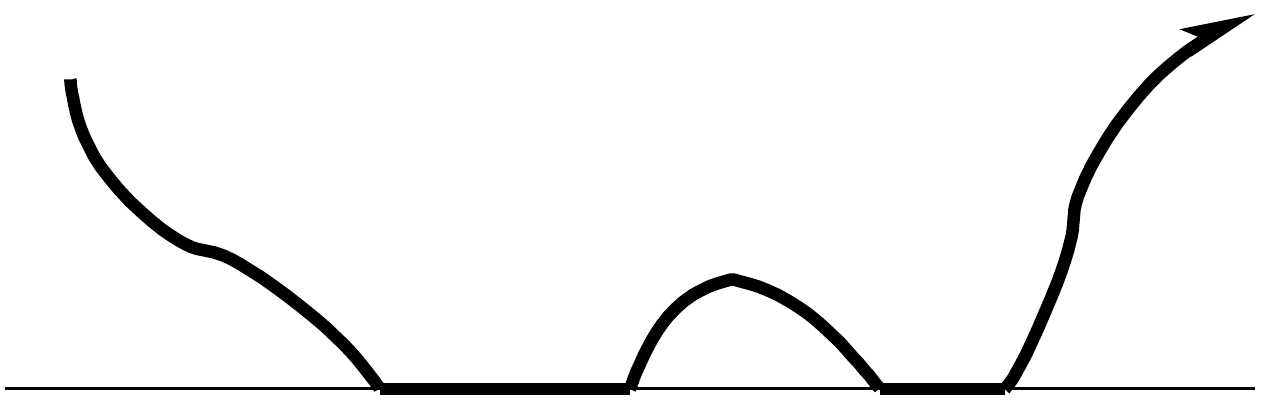}
\end{center}
\caption{The upper half plane $\Q^{(2)}$ with $\Q^{(1)}$ glued into the boundary $y=0$, as required by the ``radical topology'' of $\Q^\circ$, shown with a path of the process.}
\label{fig:radical}
\end{figure}

The radical topology may seem natural in view of \eqref{IBC4}, particularly in units in which $2mg/\hbar^2=-1$, as then the IBC is satisfied as soon as the wave function is continuous. Note, however, that the appropriate measure of volume in $\Q^\circ$ is still given by \eqref{mudef}, so the $y=0$ line (identified with $\Q^{(1)}$) is \emph{not} a null set, and $H\psi$ on $\{y=0\}$ is given by \eqref{H4def1}, not \eqref{H4def2}. It may also seem confusing that, for $Q_t$ on the boundary, there are two conflicting Bohmian equations of motion, one using $\psi^{(1)}$ and the other $\psi^{(2)}$; the way the process uses them is that for a random duration, the first equation governs the motion, and then, spontaneously at a random time $T$, the second equation takes over.
\end{enumerate}
\setcounter{remarks}{\theenumi}

\subsection{Time Reversal Symmetry}
\label{sec:reversal}

Time reversal symmetry plays a bigger role for the IBC process than for ordinary Bohmian mechanics for two reasons: first, it may be counter-intuitive that the IBC process is time reversal symmetric at all, and second, the IBC process can be characterized as the unique time reversal symmetric process in a suitable class of processes, as elucidated below.

We first describe the extension of the IBC process to negative times: The laws governing $Q_t$ are such that they define a unique process not only for all positive $t$, but also for all negative $t$; put differently, they define, for any solution $t\mapsto \psi_t$ (with $-\infty < t < \infty$) of \eqref{Schr1}, a probability distribution over paths $\RRR\ni t \mapsto Q_t \in \Q$. To see this, choose a random $|\psi_{t_0}|^2$-distributed configuration at an ``initial time'' $t_0<0$ and let the process evolve for all $t>t_0$. Since for any $t\geq t_0$, $Q_t$ will be $|\psi_t|^2$-distributed, and since it is a Markov process, the restriction of the process $(Q_t)_{t\geq t_0}$ to a time interval $[t_1,\infty)$ with $t_1>t_0$ has the same distribution as the one obtained by starting at time $t_1$. Thus, the family with parameter $t_0$ of processes $(Q_t)_{t\geq t_0}$ is consistent, and by the Kolmogorov extension theorem, each such process is the restriction to the time interval $[t_0,\infty)$ of some process $(Q_t)_{t\in\RRR}$.

Let us now turn to time reversal. Notwithstanding the fact that ``downward'' jumps (from $\Q^{(2)}$ to $\Q^{(1)}$) are deterministic (they occur when $Q_t$ hits the boundary) while ``upward'' jumps are stochastic, the process is invariant under time reversal.\footnote{Strictly speaking, to ensure reversibility, we need a reversible rule for the choice of $Q_T$ (see Remark~\ref{rem:QT}). 
Two such rules would be: (i)~$Q_T$ always lies in the lower sector (i.e., $Q_T=x$); or (ii)~$Q_T$ always lies in the higher sector (i.e., $Q_T=(x,0)$).} 
This means the following: if $\psi$ evolves according to the Hamiltonian $H=H_{IBC}$, i.e., $\psi_t=e^{-iHt/\hbar}\psi_0$, and if $(Q_t)_{t\in\RRR}$ is the associated process, then $\tilde\psi$ defined by $\tilde\psi_t=\psi_{-t}^*$ also evolves according to $H$, and $(\tilde Q_t) = (Q_{-t})$ is the process associated with $\tilde\psi$. To see this, note first that (i)~if $\psi$ satisfies the IBC then so does $\psi^*$; and (ii)~$H\psi^*=(H\psi)^*$; (i) and (ii) together imply that $\tilde\psi$ evolves according to $H$. (iii)~As is well known and obvious from \eqref{Bohm4}, the Bohmian velocity field $v^\psi=j^\psi/|\psi|^2$ changes sign when $\psi$ is replaced by $\psi^*$. Now, we need to consider the time reversal of the jumps.

\begin{figure}[h]
\begin{center}
\includegraphics[width=0.5\textwidth]{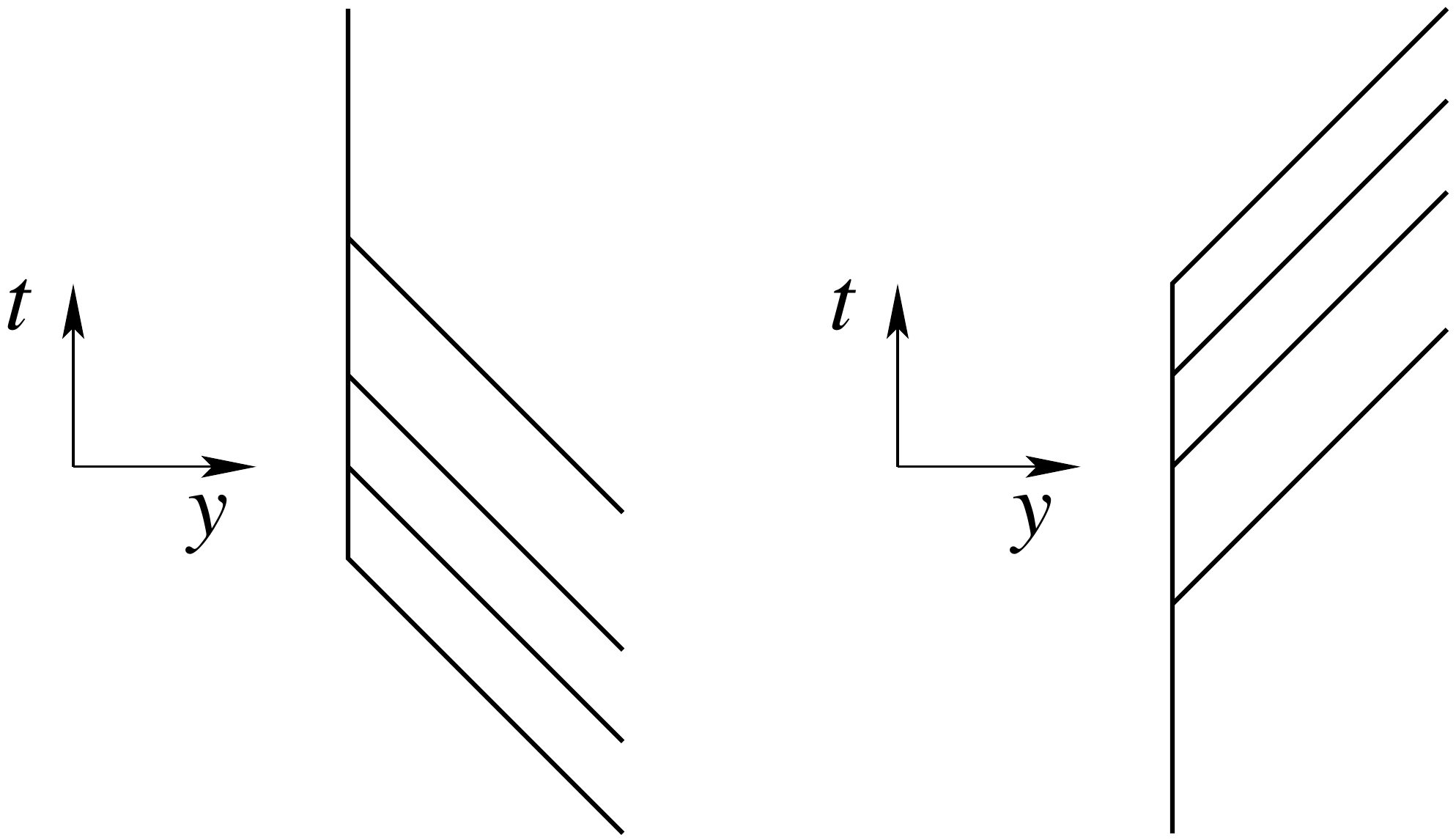}
\end{center}
\caption{Why the time reversal of a many-to-one evolution is stochastic. LEFT: Several trajectories (only $y$-component shown), depicted in a $yt$-diagram in the radical topology; the trajectories arrive at different times on the boundary and stay there on the same trajectory on the boundary. RIGHT: The time reverse of the diagram on the left; now the trajectory can leave the boundary at different times.}
\label{fig:reverse}
\end{figure}

The downward jumps erase certain information. That is, if $\tilde{x}(t)$ denotes a solution of Bohm's equation of motion \eqref{Bohm4} in $\Q^{(1)}$, then a history $(Q_t)_{0\leq t\leq t_1}$ could arrive at $Q_{t_1}=\tilde{x}(t_1)$ in various ways, including specifically a downward jump at time $\tau$ from $(\tilde{x}(\tau),0)$ to $\tilde{x}(\tau)$, followed by motion along $\tilde{x}(\cdot)$ without an upward jump. It is this many-to-one evolution that becomes, when time-reversed, stochastic (i.e., one-to-many), see Figure~\ref{fig:reverse}. The $|\psi|^2$ distribution induces a distribution over those histories ending up at $\tilde{x}(t_1)$ at time $t_1$, and this determines the distribution of the time of the upward jump in the time-reversed histories.

The mathematical criterion for the reversibility of the jumps is that the amount of probability transported by jumps from the interval $[x,x+dx]$ in $\Q^{(1)}$ to $\Q^{(2)}$ during the time interval $[t,t+dt]$ (or, so to speak, the number of histories with an upward jump from $[x,x+dx]$ during $[t,t+dt]$), given the wave function $\psi$, is equal to the amount of probability transported from the interval between $(x,0)$ and $(x+dx,0)$ in $\Q^{(2)}$ to $\Q^{(1)}$ during $[t,t+dt]$, given the wave function $\psi^*$. The former quantity is
\be\label{rev1}
\rho^{(1)}(x)\; \sigma^\psi\bigl( x \to (x,0)\bigr) \, dx\, dt\,,
\ee
the latter is 
\be\label{rev2}
\rho^{(2)}(x,0)\; \bigl[-v_y^{(2),\psi^*}(x,0)\bigr]^+ \, dx\, dt\,.
\ee
From $\rho=|\psi|^2$, the Bohmian velocity law \eqref{v2def} with \eqref{jdef}, and the jump rate law \eqref{jumprate4}, it follows that the two quantities are equal. This completes the proof of time reversal invariance.

Conversely, this reasoning determines the jump rate formula. That is, the law \eqref{jumprate4} is uniquely selected by the conjunction of the following requirements: time reversal symmetry, the Markov property, Bohm's equation of motion, deterministic jumps from $(x,0)$ to $x$, and equivariance. Indeed, consider any Markov process $Q_t$ in $\Q$ such that (i)~$Q_t$ is $|\psi_t|^2$ distributed for every $t$; (ii)~in $\Q^{(2)}$, $Q_t$ obeys Bohm's equation of motion until it hits the boundary, at which time it jumps from $(x,0)$ to $x$; (iii)~in $\Q^{(1)}$, $Q_t$ obeys Bohm's equation, except that at any time it may jump anywhere on $\partial \Q^{(2)}$.
Then, by the Markov property, the jumps occur spontaneously with some rate $\sigma_t(x\to(x',0))$; reversibility requires that the only transitions $x\to(x',0)$ that occur are the reverse of possible jumps from $\partial\Q^{(2)}$ to $\Q^{(1)}$, and thus with $x'=x$; reversibility requires further that \eqref{rev1} is equal to \eqref{rev2}, so, using $\rho=|\psi|^2$,
\be
|\psi^{(1)}(x)|^2 \; \sigma_t\bigl( x\to (x,0) \bigr) 
= |\psi^{(2)}(x,0)|^2 \, \tfrac{\hbar}{m} \Im^+ \frac{\partial_y\psi^{(2)}(x,0)}{\psi^{(2)}(x,0)}\,,
\ee
which implies \eqref{jumprate4}.

Alternatively, the jump rate formula \eqref{jumprate4} also follows without assuming time reversal symmetry if we assume instead (in addition to equivariance, the Markov property, Bohm's equation of motion, and deterministic jumps from $(x,0)$ to $x$) that the upward jumps can only be of the form $x\to (x,0)$. Indeed, if a given Bohmian trajectory in $\Q^{(2)}$ begins at $(x_0,0)\in\partial\Q^{(2)}$ at time $t_0$, then the process can reach it only by jumping to $(x_0,0)$ at time $t_0$, and it can only jump there from $x_0\in\Q^{(1)}$. For equivariance, the process's $y$-current $\rho^{(1)}(x_0,t_0) \, \sigma_{t_0}(x_0\to(x_0,0))$ out of $(x_0,0)$ at time $t_0$ must agree with $j_y^{(2)}(x_0,0,t_0)$, which implies \eqref{jumprate4} whenever $j_y^{(2)}(x,0,t)>0$; further jumps cannot occur since no trajectories begin at $(x,0)$ at time $t$ if $j_y^{(2)}(x,0,t)<0$.

\subsection{Neumann-Type and Robin-Type Boundary Conditions}

As pointed out already in \cite{TT15b} for Model~4 and in \cite{TT15a,ibc2a,IBCco1} for other models, other IBCs are possible that involve derivatives of $\psi$ normal to the boundary. While the IBC \eqref{IBC4} is of Dirichlet type in that it involves, like a Dirichlet boundary condition, the value but not the normal derivative of $\psi$ on the boundary, an IBC of Neumann type involves the normal derivative but not the value of $\psi$, and one of Robin type involves both.\footnote{However, some care is required with this terminology. For example, the ``Neumann-type'' IBC for Model~1--3, which replaces, e.g., \eqref{IBC3} by $\lim_{r\to 0} \partial_r(r\psi^{(1)}(r\vomega))=(-mg/2\pi\hbar^2)\psi^{(0)}$, has the property that for those $\psi^{(1)}$ that do not diverge at $r=0$, the left-hand side just yields $\psi^{(1)}(\vzero)$, the expression that would appear in a Dirichlet condition.} A general scheme is \cite{TT15b,IBCco1}
\begin{subequations}\label{Robin}
\begin{align}
e^{i\theta}\bigl( \alpha + \beta \partial_y\bigr)\psi^{(2)}(x,0)\:& = -\tfrac{2mg}{\hbar^2}\psi^{(1)}(x) \quad\quad \text{(IBC)}\label{RobIBC}\\[3mm]
(H\psi)^{(1)}(x)\:&=-\tfrac{\hbar^2}{2m} \partial^2_x \psi^{(1)}(x) 
+ g \, e^{i\theta}\bigl(\gamma + \delta \partial_y\bigr) \psi^{(2)}(x,0) \\[2mm]
(H\psi)^{(2)}(x,y)\:&=-\tfrac{\hbar^2}{2m} \Bigl(\partial^2_x+\partial_y^2\Bigr) \psi^{(2)}(x,y)\quad \text{for }y>0
\end{align} 
\end{subequations}
with constants $\alpha,\beta,\gamma,\delta,\theta\in \RRR$ such that
\be\label{abcd}
\alpha\delta-\beta\gamma=1\,.
\ee
(The constant $g$ can be dropped by adjusting $\alpha,\beta,\gamma,\delta$.) These equations define a self-adjoint Hamiltonian. The IBC \eqref{IBC4} and Hamiltonian \eqref{H4def} are included as the special case $\alpha=1$, $\beta=0$, $\gamma=0$, $\delta=1$, $\theta=0$.

Given constants satisfying \eqref{abcd}, a configuration process $(Q_t)$ can be defined in the same way as before, using Bohm's equation of motion again and literally the same formula \eqref{jumprate4} for the jump rate, and postulating again that the configuration, upon reaching the boundary $\partial\Q^{(2)}$ at $(x,0)$, jumps to $x\in\Q^{(1)}$. Then equivariance of $|\psi|^2$ holds again, and so does time reversal symmetry. In general, complex phases such as $e^{i\theta}$ in the coefficients of an IBC lead to violations of time reversal symmetry when the phases at different boundaries are neither equal nor opposite \cite{ST18}; this does not happen for Model~4 because it has only one boundary. However, the action of time reversal on $\psi$ is now not merely complex conjugation but involves in addition a different phase factor on each sector, viz., $\psi^{(2)}\to e^{-2i\theta}\psi^{(2)*}$ and $\psi^{(1)}\to \psi^{(1)*}$ \cite{ST18}.

\section{Particle Creation in 3 Dimensions}
\label{sec:M3M2}

We now turn to Model~3 and, later in this section, Models 2 and 1. In Model~3, a $y$-particle can move in 3-dimensional space and be absorbed and emitted by an $x$-particle fixed at the origin. We refer to the $y$-configuration simply as ``the configuration.'' Thus, the configuration space $\Q$ consists of two sectors, $\Q^{(0)}$ and $\Q^{(1)}$, corresponding to the number of $y$-particles. That is, $\Q^{(0)}$ contains only a single configuration, namely the empty configuration $\emptyset$, and $\Q^{(1)}=\RRR^3\setminus\{\vzero\}$, whose boundary $\partial\Q^{(1)}=\partial\Q=\{\vzero\}$ contains only the origin. This model lies outside the framework discussed so far because the boundary now has codimension 3. 

We will use spherical coordinates $(r,\vomega)$ in $\Q^{(1)}$ with $0< r< \infty$ and $\vomega\in\SSS^2$ (the unit sphere in $\RRR^3$), so that the boundary corresponds to $r=0$ and looks like a surface in coordinate space. In fact, it will be convenient to revise the definition of $\Q$ in the previous paragraph a little bit and set $\Q^{(1)}= [0,\infty) \times \SSS^2$ with the Riemannian metric $ds^2 = dr^2 + r^2 \, d\vomega^2$ (with $d\vomega^2$ the Riemannian metric on $\SSS^2$, so $ds^2$ becomes degenerate on the boundary $r=0$, but the IBC approach works nevertheless). With this choice of $\Q^{(1)}$, we have that $\partial\Q$ is a sphere, not a point, and that $\Q=\overline{\Q}$ in the sense of Remark~\ref{rem:Qcirc}. 
In the following, we will assume that $\psi$ is defined on $\Q^\circ$, whereas $Q_t$ moves in $\overline{\Q}$.

\subsection{Model~3}
\label{sec:M3}

The configuration space $\Q$ is equipped with the measure $\mu$ defined by
\be
\mu(S) = \begin{cases}
\mathrm{vol}(S) & \text{if }S\subseteq \Q^{(1)}\\
1+ \mathrm{vol}(S\cap \Q^{(1)}) & \text{if } \emptyset\in S\,.
\end{cases}
\ee
The Hilbert space of this model is $\Hilbert=L^2(\Q,\CCC,\mu)=\Hilbert^{(0)}\oplus \Hilbert^{(1)}$ with $\Hilbert^{(0)}=\CCC$ and $\Hilbert^{(1)}=L^2(\RRR^3,\CCC)$.
In spherical coordinates, the inner product in $\Hilbert$ reads
\be\label{inprspherical}
\scp{\psi}{\phi} = \psi^{(0)*} \phi^{(0)} + \int\limits_0^\infty \! dr \int\limits_{\SSS^2} \! d^2\vomega \: r^2 \, \psi^{(1)}(r\vomega)^* \, \phi^{(1)}(r\vomega)\,,
\ee
where $d^2\vomega$ is the surface area element on $\SSS^2$,
and the Laplace operator becomes
\be\label{Laplacespherical}
\Laplace = \partial_r^2+\tfrac{2}{r}\partial_r + \tfrac{1}{r^2} \Laplace_{\vomega}
\ee
with $\Laplace_{\vomega}$ the Laplace operator on the sphere.

The IBC demands that in Cartesian coordinates,
\be\label{IBC3}
\lim_{\vy\to\vzero} |\vy|\,\psi^{(1)}(\vy) 
= -\tfrac{mg}{2\pi\hbar^2}\; \psi^{(0)}\,,
\ee
$\vy\in \RRR^3\setminus\{\vzero\}$. Equivalently in spherical coordinates, for any sequence $r_n \to 0$ of positive numbers and any sequence $\vomega_n\in\SSS^2$,
\be\label{IBC3c}
\lim_{n\to\infty} r_n \psi^{(1)}(r_n,\vomega_n) 
= -\tfrac{mg}{2\pi\hbar^2}\; \psi^{(0)} \,.
\ee
This condition implies that, whenever $\psi^{(0)}$ is nonzero, $\psi^{(1)}(r\vomega)$ diverges as $r\to 0$ like $1/r$. The Hamiltonian is
\begin{subequations}\label{H3def}
\begin{align}
(H\psi)^{(0)} &= \tfrac{g}{4\pi} \int\limits_{\SSS^2} d^2\vomega\, \lim_{r\searrow 0} \partial_r\Bigl(r\psi^{(1)}(r\vomega)  \Bigr)\label{H3def1}\\
(H\psi)^{(1)}(r\vomega) &= -\tfrac{\hbar^2}{2m} \Bigl( \partial_r^2+\tfrac{2}{r}\partial_r + \tfrac{1}{r^2} \Laplace_{\vomega}\Bigr) \psi^{(1)}(r\vomega) \quad \text{for $r>0$}\label{H3def2}
\end{align} 
\end{subequations}
(where $r\searrow 0$ means $r\to 0$ with $r>0$).
It can be shown \cite{ibc2a} that \eqref{H3def} defines a self-adjoint operator $H$ on a dense domain $\domain$ in $\Hilbert$ consisting of functions satisfying the IBC \eqref{IBC3}.

We now define the Bell-type process $(Q_t)$ in $\overline{\Q} = \Q^{(0)} \cup \overline{\Q}^{(1)}$. If $Q_t\in\Q^{(1)}$ then it moves according to Bohm's equation of motion \eqref{Bohm4} until it hits the boundary $\{r=0\}$, at which time it jumps to $\emptyset$, where it remains for a random waiting time. The rate of jumping from $\emptyset$ to the surface element $d^2\vomega$ around $\vomega$ on the boundary $r=0$ is
\be\label{jumprate3}
\sigma_t \bigl(\emptyset\to (0,\vomega)\bigr) d^2 \vomega 
= \tfrac{\hbar}{m} \lim_{r\searrow 0}\frac{\Im^+\bigl[r^2 \psi^{(1)}(r\vomega)^* \, \partial_r \psi^{(1)}(r\vomega) \bigr]}{|\psi^{(0)}|^2} d^2\vomega\,.
\ee

The factor $r^2$, not present in the previous jump rate formula \eqref{jumprate4}, can be thought of as arising in this way: 
Since the probability current density is 
\be\label{jdef2}
\vj=\tfrac{\hbar}{m} \Im [\psi^*\nabla\psi]\,,
\ee
whose radial component is $j_r = \tfrac{\hbar}{m} \Im [\psi^*\partial_r\psi]$, the outward probability flux per time through a solid angle element $d^2\vomega$ of a sphere around the origin of radius $r$ is
\be
j_r(r\vomega)\, r^2\, d^2\vomega= \tfrac{\hbar}{m} \Im [r^2 \psi^*\partial_r\psi]\, d^2\vomega\,.
\ee
Now the outward flux from the origin in directions in the solid angle element $d^2\vomega$ is the limit thereof as $r\searrow 0$.

Let us continue the definition of $Q_t$. After jumping to $(0,\vomega)$, the process moves along the solution of Bohm's equation of motion starting at $(0,\vomega)$, see Figure~\ref{fig:spherical}. In fact, it turns out (see Remark~\ref{rem:velocity} below) that the velocity vector field $\vv^{(1)}$ in spherical coordinates possesses a continuous extension to $\overline{\Q}{}^{(1)}$, so that at most one trajectory starts at $(0,\vomega)$. As in Section~\ref{sec:M4}, the positive sign of $j_r$ at $(0,\vomega)$ guarantees that there actually is a solution of Bohm's equation beginning at $(0,\vomega)$. This completes the definition of the process $(Q_t)$.

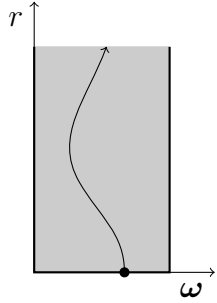
\begin{figure}[h]
\begin{center}
\begin{tikzpicture}[scale=1.2]
\draw (0,3) node[anchor=north east] {$r$};
\draw (2,0) node[anchor=north east] {$\vomega$};
\filldraw[fill=gray!40!white, draw=gray!40!white] (0,0) rectangle (1.5,2.5);
\draw[thick] (0,2.5) -- (0,0) -- (1.5,0) -- (1.5,2.5);
\draw[->] (0,0) -- (0,3);
\draw[->] (0,0) -- (2,0);
\filldraw (1,0) circle (0.5mm);
\draw[->] (1,0) to[out=90,in=300] (0.5,1) to[out=120,in=250] (0.8,2.5);
\end{tikzpicture}
\end{center}
\caption{The trajectory in $\overline{\Q}{}^{(1)}$ that $Q_t$ jumps to, represented in spherical coordinates, with only one of the two angles of $\vomega=(\varphi,\vartheta)$ drawn (shaded region = admissible values $r>0$, $0\leq \varphi<2\pi$, $0\leq \vartheta \leq \pi$). The trajectory begins at $r=0$ at a particular value of $\vomega$; the corresponding point $(0,\vomega)$ in the diagram is marked.}
\label{fig:spherical}
\end{figure}

If we want to use $\Q^\circ$ instead of $\overline{\Q}$ as the value space of $Q_t$, then, instead of ``jumping to $(0,\vomega)$ at time $T$,'' we need to say ``jumping to the trajectory whose limit backwards in time at $T$ is $\lim_{t\searrow T}Q_t=(0,\vomega)$'' and stipulate that $Q_T\in \Q^{(0)}$. If we want to use Cartesian coordinates, we need to say ``jumping to the trajectory whose limit backwards in time at $T$ is the origin ($\lim_{t\searrow T}Q_t=\vzero$) and whose limiting direction is $\vomega$ ($\lim_{t \searrow T} dQ_t/dt \propto \vomega$).'' We will often abbreviate this phrase and simply say ``jumping to $0\vomega$.''

\subsection{Equivariance}

We now derive the equivariance of $|\psi|^2$, following the lines of the argument in Section~\ref{sec:equiv} for Model~4. For $Q_t$ to be $|\psi_t|^2$ distributed means that the distribution of $Q_t$ has density $|\psi_t(q)|^2$ relative to $\mu$, i.e.,
\begin{subequations}
\begin{align}
P(Q_t=\emptyset) &= |\psi^{(0)}|^2\,,\\
P(Q_t\in d^3\vq) &= |\psi^{(1)}(\vq)|^2 \, d^3\vq
\end{align}
\end{subequations}
for a volume element $d^3\vq$ around $\vq\in\RRR^3$.
The probability transport equations for the process are
\begin{subequations}\label{transportM3}
\begin{align}
\frac{\partial \rho^{(0)}}{\partial t} 
&=\int_{\SSS^2}d^2\vomega \,\lim_{r\searrow 0} r^2 \,\rho^{(1)}(r\vomega) \, \bigl[- v^{(1)}_r(r\vomega)\bigr]^+ 
-  \rho^{(0)}\int\limits_{\SSS^2} d^2\vomega\, \sigma_t(\emptyset \to 0\vomega)\,,\\
\frac{\partial \rho^{(1)}}{\partial t} 
&= - \nabla\cdot \bigl( \rho^{(1)}\, \vv^{(1)}\bigr)\,,
\end{align}
\end{subequations}
and the equations for $|\psi|^2$ implied by the Schr\"odinger equation with Hamiltonian \eqref{H3def} are
\begin{subequations}\label{balance4}
\begin{align}
\frac{\partial |\psi^{(0)}|^2}{\partial t} 
&= \tfrac{2g}{4\pi\hbar} \Im \biggl[ \psi^{(0)*} \int\limits_{\SSS^2} d^2\vomega\, \lim_{r\searrow 0}\partial_r \Bigl(r\psi^{(1)}(r\vomega)  \Bigr)  \biggr]\,,\label{psi02}\\
\frac{\partial |\psi^{(1)}|^2}{\partial t} 
&= - \nabla\cdot \vj^{(1)}\,.
\end{align}
\end{subequations}
The IBC \eqref{IBC3} allows us to rewrite \eqref{psi02} as
\be
\frac{\partial |\psi^{(0)}|^2}{\partial t} 
= -\tfrac{\hbar}{m} \Im  \int\limits_{\SSS^2} d^2\vomega\, \lim_{r\searrow 0} \biggl[r\psi^{(1)}(r\vomega)^*\,\partial_r \Bigl(r\psi^{(1)}(r\vomega)  \Bigr)  \biggr]\,.
\ee
Using that
\be\label{Imrdr}
\Im \bigl[r\psi^*\partial_r(r\psi)\bigr] = \Im \bigl[r\psi^*\psi+r^2\psi^*\partial_r\psi\bigr] = \Im \bigl[ r^2\psi^*\partial_r\psi\bigr] \,,
\ee
we see that, by virtue of Bohm's equation \eqref{Bohm4b} and the jump rate law \eqref{jumprate3}, \eqref{balance4} agrees with \eqref{transportM3}, thus completing the derivation of equivariance.

\subsection{Remarks}

\begin{enumerate}
\setcounter{enumi}{\theremarks}

\item\label{rem:location} \textit{Location of creation.} We see that the behavior depicted in Figure~\ref{fig:radius} occurs in Model~3: The $y$-particle gets emitted \emph{at} the location of the $x$-particle (the origin). Likewise, it gets absorbed \emph{at} the location of the $x$-particle.

\item \textit{Dirichlet vs.\ Neumann vs.\ Robin conditions.} We have used a Dirichlet-type IBC for Model~3, but Neumann-type or Robin-type conditions are equally possible \cite{TT15a,ibc2a}, also with respect to the Bohmian dynamics. The version of the theory with the Dirichlet-type condition seems to be the physically most natural and relevant \cite{TT15a}.

\item\label{rem:powers} \textit{Expansion by powers of $r$.}
Let us assume for simplicity that $\psi^{(1)}$ can be expanded in powers of $r$ according to
\be\label{expansionr}
\psi^{(1)}(r\vomega) = \sum_{k=-1}^\infty r^k \, c_k(\vomega)\,.
\ee
(See Remark~\ref{rem:domain} below for a discussion of more general $\psi$ in the domain of $H$.) Note that the IBC \eqref{IBC3} enforces that an $r^{-1}$ term occurs and, at the same time, excludes any $r^\alpha$ term with $\alpha<-1$.

In terms of the coefficients $c_k$, the IBC \eqref{IBC3} can be expressed as
\be\label{IBC3b}
c_{-1}(\vomega) 
= -\tfrac{mg}{2\pi\hbar^2}\; \psi^{(0)}\,,
\ee
and the action of the Hamiltonian \eqref{H3def} as
\begin{subequations}\label{H3defb}
\begin{align}
(H\psi)^{(0)} &= \tfrac{g}{4\pi} \int\limits_{\SSS^2} d^2\vomega\, c_0(\vomega)\label{H3def1b}\\
(H\psi)^{(1)}(r\vomega) &= -\tfrac{\hbar^2}{2m} \Bigl( \partial_r^2+\tfrac{2}{r}\partial_r + \tfrac{1}{r^2} \Laplace_{\vomega}\Bigr) \psi^{(1)}(r\vomega) \quad \text{for $r>0$}\,.\label{H3def2b}
\end{align} 
\end{subequations}
From \eqref{IBC3b} it follows that $c_{-1}(\vomega)$ is actually independent of $\vomega$.

\item\label{rem:c0} \textit{$c_0$ does not depend on $\vomega$ either.} If it did, then $\Laplace \psi^{(1)}$ and thus $H\psi$ would not be square-integrable. To see this, note that
$\Laplace \psi^{(1)}$ contains contributions, arising from the third term on the right-hand side of \eqref{Laplacespherical}, of the form
\be\label{Lsexpand}
\frac{1}{r^2} \Laplace_{\vomega}\psi^{(1)} = \sum_{k=-1}^\infty r^{k-2} \, \Laplace_{\vomega}c_k(\vomega)\,.
\ee
As seen from \eqref{inprspherical}, a function of the form $r^\ell c(\vomega)$ can be square-integrable near the origin only if 
\be
\infty> \int_0^1 dr \int\limits_{\SSS^2} d^2\vomega\, r^2 \,|r^{\ell}c(\vomega)|^2 = \int_0^1 dr\, r^{2\ell+2} \int\limits_{\SSS^2}d^2\vomega\, |c(\vomega)|^2\,, 
\ee
that is, if $\ell\geq -1$. Thus, in \eqref{Lsexpand} the terms with $k=-1$ and $k=0$, if nonzero, will ruin the square-integrability (and one easily checks that this cannot be avoided by cancellations between summands of \eqref{Lsexpand}); so $\Laplace_{\vomega} c_k(\vomega)=0$ for $k=-1$ and $k=0$, which is possible on the sphere only if $c_k(\vomega)=\mathrm{const.}=c_k$ (as the eigenfunctions of $\Laplace_{\vomega}$ are the spherical harmonics, and every non-constant one of them has negative eigenvalue).

\item \textit{Unnecessary $\vomega$-integration.}\label{rem:integration} As a curious consequence of the previous remark, we can actually drop the $\vomega$-integration in the definition \eqref{H3def1} of $H$ (along with the prefactor $1/4\pi$) because, for $\psi$ in the domain, the integrand is $c_0$ and thus $\vomega$-independent. In fact, \eqref{H3def1} is equivalent to $(H\psi)^{(0)}= g\, c_0$.

\item \textit{Uniformity over the sphere.}\label{rem:uniform} It turns out that the jump rate \eqref{jumprate3} does not, in fact, depend on $\vomega$, so that the jump destination $(0,\vomega)$ is always chosen with uniform distribution over the sphere, and only the rate of jumping at all depends on $\psi$,
\be
\sigma_t (\emptyset\to \Q^{(1)}) 
= \tfrac{4\pi\hbar}{m} \lim_{r\searrow 0}\frac{\Im^+\bigl[r^2 \psi^{(1)}(r\vomega)^* \, \partial_r \psi^{(1)}(r\vomega) \bigr]}{|\psi^{(0)}|^2} \quad\quad \forall \vomega\in\SSS^2.
\ee

This follows from the fact that the radial current $j_r$ at $r=0$ does not depend on $\vomega$; in fact,
\be\label{r2jr}
\lim_{r\searrow 0} r^2j_r^{(1)}(r\vomega) = \tfrac{\hbar}{m} \Im[c_{-1}^*c_0]\,,
\ee
so that, in particular, the jump rate $\sigma_t$ can equivalently be written in the form
\be\label{jumprate3b}
\sigma_t \bigl( \emptyset \to (0,\vomega) \bigr) = \tfrac{\hbar}{m} \frac{\Im^+ [c_{-1}^* c_0]}{|\psi^{(0)}|^2}\,,
\ee
independently of $\vomega$. It follows also from \eqref{r2jr} that the current out of the origin is
\be
J_0 := \lim_{r \searrow 0} r^2 \int_{\SSS^2} \!\! d\vomega \, j_r^{(1)}(r\vomega) = \tfrac{4\pi\hbar}{m} \Im[c_{-1}^* c_0]\,.
\ee

In order to derive \eqref{r2jr}, we begin with \eqref{Imrdr}. As a consequence, the limit does not necessarily commute with taking the imaginary part, 
\be
\lim_{r\searrow 0} \Im[r^2 \psi^* \partial_r \psi] \neq \Im \lim_{r\searrow 0} [r^2\psi^*\partial_r\psi]
\ee
(the latter limit need not exist, as the real part may diverge); however,
\begin{align}
\lim_{r\searrow 0} r^2 j_r^{(1)}(r\vomega) 
&= \tfrac{\hbar}{m} \lim_{r\searrow 0} \Im[r^2\psi^*\partial_r\psi]\\
&= \tfrac{\hbar}{m}\lim_{r\searrow 0} \Im[r\psi^* \partial_r(r\psi)]\\ 
&= \tfrac{\hbar}{m}\Im \lim_{r\searrow 0}[r\psi^*\partial_r(r\psi)]\,.
\end{align}
From the expansion \eqref{expansionr}, we have that $r\psi = c_{-1} + c_0 r + O(r^2)$ and $\partial_r(r\psi) = c_0 + O(r)$, so $r\psi^* \partial_r(r\psi) = c_{-1}^* c_0 + O(r)$, which yields \eqref{r2jr}.

\item \textit{Velocity of emitted particle.}\label{rem:velocity} As another consequence of Remark~\ref{rem:uniform}, the initial velocity of a newly emitted $y$-particle always points radially in spherical coordinates, see Figure~\ref{fig:radial1}, and its magnitude is independent of $\vomega$. 

\begin{figure}[h]
\begin{center}
\begin{tikzpicture}[scale=1.2]
\draw (0,3) node[anchor=north east] {$r$};
\draw (2,0) node[anchor=north east] {$\vomega$};
\filldraw[fill=gray!40!white, draw=gray!40!white] (0,0) rectangle (1.5,2.5);
\draw[thick] (0,2.5) -- (0,0) -- (1.5,0) -- (1.5,2.5);
\draw[->] (0,0) -- (0,3);
\draw[->] (0,0) -- (2,0);
\filldraw (1,0) circle (0.5mm);
\draw[->] (1,0) to[out=90,in=300] (0.5,1) to[out=120,in=250] (0.8,2.5);
\draw (4,3) node[anchor=north east] {$r$};
\draw (6,0) node[anchor=north east] {$\vomega$};
\filldraw[fill=gray!40!white, draw=gray!40!white] (4,0) rectangle (5.5,2.5);
\draw[thick] (4,2.5) -- (4,0) -- (5.5,0) -- (5.5,2.5);
\draw[->] (4,0) -- (4,3);
\draw[->] (4,0) -- (6,0);
\filldraw (5,0) circle (0.5mm);
\draw[->] (5,0) to[out=45,in=300] (4.5,1.5) to[out=120,in=250] (4.8,2.5);
\end{tikzpicture}
\end{center}
\caption{Different kinds of trajectories, represented in spherical coordinates, one (LEFT) with $v_{\vomega}=0$ at $r=0$ and one (RIGHT) with $v_{\vomega}\neq 0$ at $r=0$. The latter case does not occur in the Bohmian dynamics we are discussing.}
\label{fig:radial1}
\end{figure}
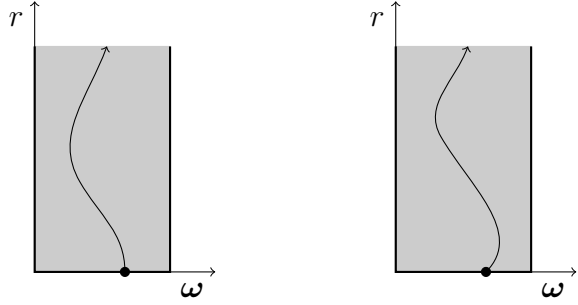

To see this, note that the radial component of the Bohmian velocity vector field at $r=0$ is, by virtue of \eqref{r2jr},
\be\label{vr}
\lim_{r\searrow 0} v_r^{(1)}(r\vomega) 
= \lim_{r\searrow 0} \frac{r^2 \, j_r^{(1)}(r\vomega)}{|r \, \psi^{(1)}(r\vomega)|^2} 
= \tfrac{\hbar}{m} \frac{\Im [c_{-1}^*\,c_0]}{|c_{-1}|^2}\,,
\ee
and thus $\vomega$-independent, while its angular components are
\be\label{vomega}
\lim_{r\searrow 0} v_{\vomega}^{(1)}(r\vomega) 
= \lim_{r\searrow 0} \frac{r^2 \, j_{\vomega}^{(1)}(r\vomega)}{|r \, \psi^{(1)}(r\vomega)|^2} 
= \tfrac{\hbar}{m} \frac{\Im [c_{-1}^*\nabla_{\!\vomega}c_{-1}]}{|c_{-1}|^2}=0\,.
\ee
Of course, as soon as $r>0$, the velocity can change, and $v_{\vomega}$ need no longer vanish. By the way, even if $v_{\vomega}\neq 0$ at $r=0$, the velocity vector \emph{in Cartesian coordinates} is still pointing radially outward, in fact in the direction $\vomega$ if the trajectory starts at $(0,\vomega)$, see Figure~\ref{fig:radial2}. (Readers might think that curves with $v_{\vomega}=0$ have curvature 0 at $r=0$, but this is not true of all curves; a counterexample is given by $d\varphi/dr=r\sin(1/r)$, whose curvature does not approach 0 as $r\to 0$.)

\begin{figure}[h]
\begin{center}
\begin{tikzpicture}[scale=4]
\draw[->] (-0.2,0) -- (0.4,0);
\draw[->] (0,-0.2) -- (0,0.9);
\draw[thick,domain=0:100,samples=200] plot ({sqrt(\x)*cos(\x)/15},{sqrt(\x)*sin(\x)/15}); 
\draw[->] (0.8,0) -- (1.4,0);
\draw[->] (1,-0.2) -- (1,0.9);
\draw[thick,domain=0:100,samples=100] plot ({1+\x*cos(\x)/150},{\x*sin(\x)/150}); 
\end{tikzpicture}
\end{center}
\caption{Examples of curves (shown in Cartesian coordinates) that have (LEFT) $v_{\vomega}=0$ at $r=0$ and (RIGHT) $v_{\vomega}\neq 0$ at $r=0$. Both have $\vomega$ pointing to the right at $r=0$, and correspondingly start from the origin to the right.}
\label{fig:radial2}
\end{figure}
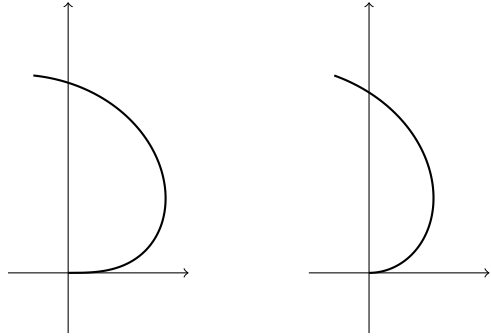

\item\label{rem:domain} \textit{Expansion in the general case.} The functions in the domain of $H$ satisfy the IBC \eqref{IBC3}, but not all of them are analytic, so not all of them can be expanded in a power series as in \eqref{expansionr}. However \cite{ibc2a}, all of them can be expanded to a certain extent, according to
\be\label{expand1}
\psi^{(1)}(r\vomega) = c_{-1} r^{-1} + c_0 r^0 + \varepsilon(r\vomega)\,,
\ee
where $c_{-1}$ and $c_0$ are complex coefficients independent of $\vomega$, and $\varepsilon(\cdot):\RRR^3\to\CCC$ is a function from the second Sobolev space $H^2(\RRR^3,\CCC)$ with $\varepsilon(\vzero)=0$. (By the Sobolev imbedding theorem, all functions in $H^2(\RRR^3,\CCC)$ are continuous.) It follows that
\be
\lim_{r\searrow 0}  \int\limits_{\SSS^2} d^2\vomega \Bigl| \partial_r \varepsilon(r\vomega) \Bigr|<\infty \,,
\ee
and from this one can obtain the same conclusions as in Remarks~\ref{rem:powers}--\ref{rem:velocity} above.

\item \textit{$1/r^2$ divergence of probability density.}\label{rem:1/r^2} From \eqref{vr} and \eqref{vomega} one can also draw conclusions about the velocity right before absorption: its direction is radial in spherical coordinates, and its magnitude is independent of $\vomega$. In particular, the velocity field is asymptotically, as $\vy\to 0$, of the form
\be\label{vasymptotic}
\vv^{(1)}(\vy) = -\alpha \frac{\vy}{|\vy|} + o(1)
\ee
with constant $\alpha= -(\hbar/m)\Im [c_0/c_{-1}]$, which is positive in the case of absorption. The motion according to the equation
\be
\frac{d\vy}{dt} = -\alpha \frac{\vy}{|\vy|}\,,
\ee
i.e., radial inward motion at constant speed $\alpha$, has the property of transporting a radial shell of radius $r_1$ and thickness $dr$ (which has volume $4\pi r_1^2dr$) to a shell of radius $r_2<r_1$ and thickness $dr$ (which has volume $4\pi r_2^2dr$) within time $t=(r_1-r_2)/\alpha$. Now suppose a radially symmetric probability distribution gets transported; since the volume of the shell goes down like $r^2(t)$ while the amount of probability stays constant, probability density will increase like $1/r^2(t)$. In particular, the stationary probability density is proportional to $1/r^2$. This fits with the Born rule $\rho=|\psi|^2$ and the situation that $\psi^{(1)}$ diverges at the origin like $1/r$. In fact, if the asymptotics \eqref{vasymptotic} of the velocity field are assumed, then we are forced to allow $\psi$ to diverge like $1/r$. Of course, the same reasoning could be done with emission instead of absorption.

\item \textit{Different rate of divergence in 2 dimensions.}\label{rem:logr} The reasoning of the previous remark might suggest that if we considered the same model in 2 space dimensions, then probability density should diverge like $1/r$ as $r\to 0$, and thus that $\psi$ should diverge like $1/\sqrt{r}$. This is actually not the case. Instead \cite{LS18}, in 2 dimensions, the expansion \eqref{expand1} gets replaced by
\be\label{expand12}
\psi^{(1)}(r\vomega) = c_\ell \log r + c_0 r^0 + \varepsilon(r\vomega)
\ee
for $\vomega \in \SSS^1$, so that the probabiliy density actually diverges like $\log^2 r$. This behavior occurs together with the fact that \eqref{vasymptotic} is not valid in 2d, as the limiting radial velocity is infinite. In fact, instead of \eqref{vasymptotic} the asymptotic velocity reads
\be\label{vasymptotic2}
\vv^{(1)}(\vy) = -\alpha \frac{\vy}{|\vy|^2 \log^2 |\vy|} +O\Bigl(\frac{1}{\log |\vy|}\Bigr)
\ee
with constant $\alpha = (\hbar/m)\, \Im[c_0/c_\ell]$, provided that $c_\ell\neq 0$. And for the motion
\be
\frac{d\vy}{dt} = -\alpha \frac{\vy}{|\vy|^2 \log^2 |\vy|} \,,
\ee
the stationary radially symmetric density is indeed proportional to $\log^2 r$. The condition $c_\ell\neq 0$ is required for a nonzero current into (or out of) the origin, as the current out of the origin is 
\be
J_0 = \lim_{r \searrow 0} r\int_{\SSS^1}d\vomega \, j_r^{(1)}(r\vomega) = \tfrac{2\pi\hbar}{m} \Im[c_0^* c_\ell]\,.
\ee

\item \textit{Another type of UV cut-off.}\label{rem:deltasphere} A UV cut-off is usually implemented by either discretizing space or smearing out the electron over a ball with (say) radius $\delta>0$ with some profile function $\varphi$ that can be regarded as the charge density of the electron and that replaces the Dirac delta function in the Hamiltonian (see, e.g., \cite{TT15a}). Another kind of UV cut-off \cite{IBCco1} consists of smearing out the electron charge not over a ball but over a sphere; correspondingly, a $y$-particle gets absorbed by an $x$-particle as soon as their distance gets as small as $\delta$; this amounts to introducing a boundary in configuration space. This cut-off was, as far as we know, first described (and implemented by means of an IBC) in \cite{TG04}. In the context of Model~3, this cut-off amounts to putting the boundary of $\Q^{(1)}$ at $r=\delta$ instead of $r=0$, and the equations of Model~3 can be adapted straightforwardly with $(\delta,\vomega)$ replacing $(0,\vomega)$; Remarks~\ref{rem:location} and \ref{rem:powers}--\ref{rem:logr} then no longer apply. An adapted version of the proof from \cite{ibc2a} shows that the Hamiltonian with this UV cut-off is well defined and self-adjoint. Since the boundary then has codimension 1, the resulting model is a special case of the framework described in Section~\ref{sec:co1}. A basic fact and crucial realization (more recent than \cite{TG04}) about IBCs is that this cut-off can be removed by taking the limit $\delta\to0$, or rather, that no cut-off need be introduced as we can set $\delta=0$ to begin with (as we have done in Section~\ref{sec:M3}); that is, that IBCs provide a UV-finite theory without UV cut-off. It seems plausible to conjecture that in the limit $\delta\to0$, the IBC process with cut-off on the $\delta$-sphere converges to the IBC process without cut-off.
\end{enumerate}
\setcounter{remarks}{\theenumi}

\subsection{Model~2}
\label{sec:M2}

In Model~2, whose IBC and Hamiltonian we first described in \cite{TT15a}, we drop the limitation of the number of $y$-particles to 0 and 1 (while there is still only one $x$-particle fixed at the origin). As the Hilbert space, we use the bosonic Fock space,
\be
\Hilbert = \bigoplus_{n=0}^\infty \Sym L^2(\RRR^3,\CCC)^{\otimes n}\,,
\ee
where $\Sym$ is the symmetrization operator and $\Sym L^2(\RRR^3,\CCC)^{\otimes n}$ its range, the space of permutation-symmetric functions of $n$ arguments in $\RRR^3$. As the configuration space, we may either choose the space of \emph{ordered} configurations,
\be
\Q^\circ=\bigcup_{n=0}^\infty (\RRR^3\setminus \{\vzero\})^n =: \bigcup_{n=0}^\infty \Q^{(n)}\,,
\ee
or the space of \emph{unordered} configurations (see, e.g., \cite{DGTTZ06,fermionic}),
\be
\tQ{}^\circ=\Bigl\{q\subset \RRR^3\setminus\{\vzero\}: \#q<\infty\Bigr\}\,.
\ee
The latter can be obtained from the former by removing configurations with two or more $y$-particles at the same location and identifying configurations that differ merely by a permutation. While $\tQ{}^\circ$ is physically more reasonable, we will use $\Q^\circ$ because it is somewhat simpler and more familiar. Since in this model the particle reactions (i.e., creation or annihilation events) occur when an $x$ and a $y$ meet, but not when two $y$'s meet, it does not matter whether we remove configurations with two $y$'s at the same location, and the relevant boundary $\partial \Q^{(n)}$ consists of configurations with a $y$-particle at the origin. Fock vectors $\psi\in\Hilbert$ can be regarded as functions $\psi:\Q^\circ\to\CCC$ such that each $\psi^{(n)}$ (the restriction of $\psi$ to $\Q^{(n)}$) is permutation-symmetric. 

The asymptotics of wave functions in the domain of $H$ near the boundary are
\be\label{expand2}
\psi(y) = c_{-1}(y\setminus \vy_j) \, r_j^{-1}+ c_0(y\setminus \vy_j) \, r_j^0 + o(r_j^0)
\ee
as $r_j:=|\vy_j|\searrow 0$ for $y=(\vy_1,\ldots,\vy_n)\in(\RRR^3\setminus\{\vzero\})^n$ and any $n\geq 1$, using the notation $y\setminus \vy_j = (\vy_1,\ldots,\vy_{j-1},\vy_{j+1},\ldots,\vy_n)$. By permutation symmetry, the functions $c_{-1}$ and $c_0$ do not depend on the choice of $j$ (while they do depend on $n$).

The IBC asserts that for every $n>0$ and every $j\leq n$,
\be\label{IBC2}
\lim_{\vy_j\to \vzero} \,|\vy_j| \,\psi(y)
= -\tfrac{mg}{2\pi\hbar^2\sqrt{n}}\, \psi(y\setminus \vy_j)\,.
\ee
The corresponding Hamiltonian is 
\begin{align}
H\psi(y) 
&=-\tfrac{\hbar^2}{2m} \sum_{j=1}^{n} \nabla^2_{\vy_j}\psi(y)+ nE_0 \psi(y) \nonumber\\
& + \tfrac{g\sqrt{n+1}}{4\pi}\int\limits_{\SSS^2} \!\! d^2\vomega \, \lim_{r\searrow 0} \partial_r \Bigl[ r \psi\bigl(y,r\vomega \bigr) \Bigr]\nonumber\\
& +\: \tfrac{g}{\sqrt{n}} \sum_{j=1}^n  \delta^3(\vy_j)\,\psi\bigl(y\setminus \vy_j\bigr)\,,\label{H2def}
\end{align}
where we have introduced a constant $E_0\geq 0$ that represents the amount of energy that must be expended to create a $y$-particle. (For $n=0$, the first and the last line should be understood as vanishing.)
It has been shown \cite{ibc2a} that on a certain dense subspace $\domain$ of $\Hilbert$, the elements of which satisfy the IBC \eqref{IBC2}, the operator $H$ given by \eqref{H2def} is well-defined, self-adjoint, and bounded from below. 

A feature of the expression \eqref{H2def} requires further explanation: the Dirac $\delta^3$ factor in the last line. Such a factor would make a Hamiltonian without IBC UV divergent, but causes no problem here for the following reason. Wave functions satisfying the IBC \eqref{IBC2}, which diverge like $1/|\vy_j|$ as $\vy_j\to \vzero$, contribute Dirac delta functions to $\nabla_{\vy_j}^2 \psi$; viz., since
\be
\nabla_{\vx}^2 \frac{1}{|\vx|} = -4\pi \delta^3(\vx)\,,
\ee
we have that
\begin{align}
-\tfrac{\hbar^2}{2m} \nabla_{\vy_j}^2 \psi(y)
&= \tfrac{\hbar^2}{2m} \nabla_{\vy_j}^2 \tfrac{mg}{2\pi\hbar^2\sqrt{n}} \frac{\psi(y\setminus \vy_j)}{|\vy_j|} + \text{a function}\\
&= -\tfrac{g}{\sqrt{n}} \delta^3(\vy_j)\, \psi(y\setminus \vy_j) + \text{a function.} 
\end{align}
These contributions cancel the last line of \eqref{H2def}. 

Why did no delta function appear in the formula \eqref{H3def2} for the Hamiltonian of Model~3? Because that formula was stated only for $r>0$, where the delta function does not contribute. In fact, it would be correct to re-write \eqref{H3def2} as
\be
(H\psi)^{(1)}(\vy)
=-\tfrac{\hbar^2}{2m}\nabla_{\vy}^2\psi^{(1)}(\vy) + g \,\delta^3(\vy)\,\psi^{(0)}\,.
\ee

We now define the process $(Q_t)$ in $\overline{\Q}$ analogously to that of Model~3. If $Q_t=y=(\vy_1,\ldots,\vy_n)\in\Q^{(n)}$ with $n\geq 0$, then with jump rate
\be\label{jumprate2}
\sigma_t \;  d^2 \vomega 
= \tfrac{\hbar}{m} \lim_{r\searrow 0}\frac{\Im^+\bigl[r^2 \psi(y,r\vomega)^* \, \partial_r \psi(y,r\vomega) \bigr]}{|\psi(y)|^2} d^2\vomega
\ee
it jumps to the solution of Bohm's equation in $\Q^{(n+1)}$ beginning at
\be
(\vy_1, \ldots, \vy_{j-1}, 0\vomega, \vy_j,\ldots, \vy_n)
\ee
with $1\leq j\leq n+1$. That is, the newly created $y$-particle gets inserted at the $j$-th position, where $j$ is chosen uniformly random. (It does not matter whether \eqref{jumprate2} involves $\psi(\vy_1, \ldots, \vy_{j-1}, r\vomega, \vy_j,\ldots\vy_n)$ or $\psi(y,r\vomega)$ because of the permutation symmetry of $\psi$.) Again, the right-hand side of \eqref{jumprate2} is actually independent of $\vomega$, so $\vomega$ is also chosen random with uniform distribution (see Remark~\ref{rem:uniform} in Section~\ref{sec:M3}).

As long as $Q_t$ does not jump to the next higher sector, it follows the solution of Bohm's equation of motion \eqref{Bohm4}, now understood as applying in $\Q^{(n)}$, until it hits the boundary, i.e., one of the $y$-particles (say, $\vy_j$) reaches the origin, in which event the process jumps to $y\setminus\vy_j$. This completes the definition of the process. 

Equivariance can be established in the same manner as before. 

Keppeler and Sieber \cite{KS16} have proposed an IBC Hamiltonian for particle creation in 1 dimension. A process for this Hamiltonian can be set up in an analogous way.
An explicit example of the IBC process for a particular wave function in a variant of Model~2 with two sources is described in Section~2.6 of \cite{ST18}.

\bigskip

\noindent{\bf Remark.}
\begin{enumerate}
\setcounter{enumi}{\theremarks}
\item \textit{Ground state.} The ground state of Model~2 can be computed explicitly \cite{TT15a,ibc2a}. It is a superposition of contributions from all sectors, and its $n$-particle wave function $\psi^{(n)}$ is, up to an $n$-dependent constant, a product of $n$ copies of a 1-particle wave function. Since this wave function is real up to a global phase factor, nothing moves. That is, the Bohmian velocities and the jump rates are zero. This behavior occurs as well in the known Bell-type QFTs with UV cut-off \cite{DGTZ03}, \cite[Section 6.4]{DGTZ05a}. It can be regarded as a consequence of the time-reversal symmetry; in fact, for any $H$ that commutes with complex conjugation and any non-degenerate eigenstate $\psi$ of $H$ (that then can be taken as real), the time-reversed wave function coincides with $\psi$ itself, so all currents vanish, and $Q_t$ is time-independent. Correspondingly, this behavior usually no longer occurs if time-reversal symmetry fails, which happens \cite{ST18} if the coefficient $g$ (the ``charge'') is made complex and given different values for different $x$-particles, in such a way that their phases are neither equal nor opposite.
\end{enumerate}
\setcounter{remarks}{\theenumi}

\subsection{Model~1: Moving Sources}
\label{sec:M1}

To obtain a full QFT, we now turn to Model~1 and allow the $x$-particles to move as well. The IBC and Hamiltonian were discussed in \cite{TT15a,Lam18}. The configuration space is 
\be
\Q = \bigcup_{m=0}^\infty \bigcup_{n=0}^\infty\Q^{(m,n)} = \bigcup_{m=0}^\infty \bigcup_{n=0}^\infty \RRR^{3m}_x \times \RRR^{3n}_y\,,
\ee
with the boundary $\partial\Q$ formed by those configurations with a $y$-particle at the same location as an $x$-particle, except that, as in Model~2, it will be convenient to treat $r=0$ as a sphere. The Hilbert space is $\Hilbert=\Hilbert_x \otimes \Hilbert_y$, with $\Hilbert_x$ and $\Hilbert_y$ the fermionic and bosonic Fock spaces, respectively,
\begin{subequations}
\begin{align}
\Hilbert_x &= \bigoplus_{m=0}^\infty \Anti L^2(\RRR^3,\CCC)^{\otimes m} \\
\Hilbert_y &= \bigoplus_{n=0}^\infty \Sym L^2(\RRR^3,\CCC)^{\otimes n}\,,
\end{align}
\end{subequations}
where $\Anti$ is the anti-symmetrization operator. Let $m_x,m_y>0$ be the mass of an $x$-particle and a $y$-particle, respectively.

The asymptotics of wave functions $\psi:\Q^\circ\to \CCC$ in the domain of $H$ near the boundary surface $\{\vx_i=\vy_j\}$ in $\Q^{(m,n)}= \RRR_x^{3m} \times \RRR_y^{3n}$ are \cite{Lam18}
\be\label{expand3}
\psi(x,y) = c_{-1,i}(x,y\setminus \vy_j) \, r_{ij}^{-1} + c_{\ell,i} (x,y\setminus \vy_j) \, \log r_{ij} + c_{0,i} (x,y\setminus \vy_j) + o(r_{ij}^0)\,,
\ee
where $r_{ij}=|\vx_i-\vy_j|$, $x=(\vx_1,\ldots,\vx_m)$, and $y=(\vy_1,\ldots,\vy_n)$ with $\vx_k\neq \vy_r$ for all $k,r$.
By permutation symmetry, the functions $c_{-1,i}, c_{\ell,i}$, and $c_{0,i}$ do not depend on the choice of $j$ (while they do depend on $m$ and $n$). Moreover, in order to make $H$ self-adjoint, the coefficients are related according to \cite{Lam18}
\be\label{cell}
c_{\ell,i} = \eta\, c_{-1,i}
\ee
with real proportionality factor
\be
\eta = \frac{m_x^2}{2\pi^2(m_x+m_y)^2} \Biggl[\frac{\sqrt{m_x(m_x+2m_y)}}{m_x+m_y}-\frac{m_x+m_y}{m_y}\arctan \biggl( \frac{m_y}{\sqrt{m_x(m_x+2m_y)}}  \biggr)  \Biggr]\,.
\ee
The IBC demands that for 
any $i=1,\ldots, m$ and any $j=1,\ldots, n$,
\be\label{IBC1}
\lim_{(\vx_i,\vy_j)\to(\vx,\vx)} \,r_{ij} \,\psi(x,y)
= -\tfrac{\mu g}{2\pi\hbar^2\sqrt{n}}\, \psi\bigl(\vx_i=\vx,\widehat{\vy_j}\bigr)\,,
\ee
where $\widehat{\ }$ denotes omission
and $\mu$ is the relative mass (or reduced mass),
\be\label{mumassdef}
\mu= \frac{m_xm_y}{m_x+m_y}\,.
\ee
Here, $\psi(\vx_i=\vx,\widehat{\vy_j})$ means $\psi(\vx_1,\ldots,\vx_{i-1},\vx,\vx_{i+1},\ldots,\vx_m,\vy_1,\ldots,\vy_{j-1},\vy_{j+1},\ldots,\vy_n)$. Equivalently, the IBC can be expressed as
\be\label{IBC1b}
c_{-1,i}(x,y) = -\tfrac{\mu g}{2\pi\hbar^2\sqrt{n}} \, \psi(x,y)\,.
\ee
In particular, $c_{-1,i}$ (and thus also $c_{\ell,i}$) actually does not depend on $i$. The Hamiltonian is
\begin{align}
&(H\psi)(x,y) 
= -\tfrac{\hbar^2}{2m_x} \sum_{i=1}^{m} \nabla^2_{\vx_i}\psi(x,y)
-\tfrac{\hbar^2}{2m_y} \sum_{j=1}^{n} \nabla^2_{\vy_j}\psi(x,y)+ nE_0 \psi(x,y) \nonumber\\
& ~~~~~~~~~~~ + g\sqrt{n+1}\sum_{i=1}^{m} c_{0,i}(x,y) 
+ \tfrac{g}{\sqrt{n}} \sum_{i=1}^m \sum_{j=1}^n  \delta^3(\vx_i-\vy_j)\,\psi\bigl(x,y\setminus \vy_j\bigr)\,. \label{H1def}
\end{align}
Again, this Hamiltonian is well-defined and self-adjoint \cite{Lam18}. Since $H$ contains no terms for the creation or annihilation of $x$-particles, it commutes with the $x$-particle number operator; in the Bell-type process, as we will see, the actual $x$-particle number is conserved. (This entails that parts of the wave function in sectors with $x$-particle number different from the actual number play no role for the particle trajectories, so that a ``strong superselection rule'' \cite{CDT05} holds for $x$-particle number.)

The process $Q_t\in\overline{\Q}$ is defined as follows. Between jumps, it follows Bohm's equation of motion, which now reads
\begin{subequations}\label{Bohm1}
\begin{align}
\label{Bohmx}
\frac{d\vX_i}{dt} &= \tfrac{\hbar}{m_x} \, \Im \, \frac{\nabla_{\vx_i}\psi}{\psi}(Q_t)\,,\\
\label{Bohmy}
\frac{d\vY_j}{dt} &= \tfrac{\hbar}{m_y} \,\Im\, \frac{\nabla_{\vy_j}\psi}{\psi}(Q_t)
\end{align}
\end{subequations}
for $Q_t=(\vX_1,\ldots,\vX_m,\vY_1,\ldots,\vY_n)$. As soon as a $y$-particle reaches an $x$-particle, it gets removed from the configuration. When at $(x,y)=(\vx_1,\ldots,\vx_m,\vy_1,\ldots,\vy_n)$, the process spontaneously jumps to the solution of Bohm's equation in $\Q^{(m,n+1)}$ starting at
\be
\Bigl( \vx_1,\ldots,\vx_{i-1},\vx_i-0\vomega,\vx_{i+1},\ldots,\vx_m, \vy_1,\ldots, \vy_{j-1}, \vx_i + 0\vomega,\vy_{j},\ldots,\vy_n \Bigr)
\ee
with rate
\be\label{jumprate1}
\sigma_t \;  d^2 \vomega 
= \tfrac{\hbar}{\mu} \lim_{r\searrow 0}\frac{\Im^+\bigl[r^2 \psi\bigl(\mathrm{cr}(x,y;i,r,\vomega)\bigr)^* \, \partial_r \psi\bigl(\mathrm{cr}(x,y;i,r,\vomega)\bigr) \bigr]}{|\psi(x,y)|^2} d^2\vomega
\ee
with $\mu$ as in \eqref{mumassdef} and
\be
\mathrm{cr}(x,y;i,r,\vomega) = \Bigl( \vx_1,\ldots,\vx_{i-1},\vx_i-\tfrac{\mu}{m_x} r\vomega,\vx_{i+1},\ldots,\vx_m,\vy_1,\ldots,\vy_n,\vx_i+\tfrac{\mu}{m_y} r\vomega\Bigr)
\ee
in $\Q^{(m,n+1)}$. 
The rate \eqref{jumprate1} is again $\vomega$-independent as in Remark~\ref{rem:uniform} in Section~\ref{sec:M3} and can equivalently be written in the form
\begin{align}
\sigma_t &= \tfrac{\hbar}{\mu} \frac{\Im^+ [c_{-1}^*(x,y) \, c_{0,i}(x,y)]}{|\psi(x,y)|^2} \label{jumprate1b}\\
\text{or}~~\sigma_t &= \tfrac{g}{2\pi\hbar\sqrt{n}} \Im^+ [- c_{0,i}(x,y) / \psi(x,y)]  \label{jumprate1c}\\
\text{or}~~\sigma_t &= \tfrac{\mu g^2}{4\pi^2\hbar^3 n }\Im^+ [c_{0,i}(x,y)/c_{-1}(x,y)]\,. \label{jumprate1d}
\end{align}
Indeed, by \eqref{expand3} and a reasoning analogous to that of Remark~\ref{rem:uniform}, at $\mathrm{cr}(x,y;i,r,\vomega)$,
\begin{align}
\psi^* \partial_r \psi 
&= \Bigl(c_{-1}^* r^{-1} + c_{\ell}^* \log r + c_0^* + o(1)\Bigr) \Bigl(-c_{-1} r^{-2} + c_{\ell}r^{-1} + o(r^{-1})\Bigr)\\
&=-|c_{-1}|^2 r^{-3} -(c_\ell^*c_{-1}) r^{-2} \log r + (c_{-1}^* c_\ell -c_0^* c_{-1})r^{-2} + o(r^{-2})\,.
\end{align}
The $r^{-3}$ term has vanishing imaginary part, and so do, by virtue of \eqref{cell}, the $r^{-2}\log r$ term and the first contribution to the $r^{-2}$ term. As a consequence, 
\be
\Im [r^2 \psi^* \partial_r \psi] = \Im [-c_0^* c_{-1}] + o(1)\,,
\ee
which yields \eqref{jumprate1b}; application of the IBC \eqref{IBC1b} then yields \eqref{jumprate1c} and \eqref{jumprate1d}.

Equivariance can be checked in the same manner as before. As in Remark~\ref{rem:velocity} it follows that the velocity of emission (or absorption) is radial in spherical coordinates, and its magnitude is finite and independent of $\vomega$.

\subsection{Renormalization}
\label{sec:renormalize} 

For our purposes, renormalization means to consider a QFT with UV cut-off and take the limit in which the cut-off is removed, if such a limit exists. This leads to the three questions whether the limiting Hamiltonian agrees with the IBC Hamiltonian (answered in the positive for Model~2 in \cite{ibc2a}), whether the Bell-type process defined in \cite{Bell86,DGTZ03,DGTZ04,DGTZ05a,DGTZ05b} (see also Section~\ref{sec:Bell-type} below) for the cut-off theories possesses a limit, and whether this limit agrees with the IBC process defined here.

One way of implementing a UV cut-off is to discretize space, and in Section~\ref{sec:lattice} below we will show non-rigorously for Model~4 as an example that, at least if the discrete Hamiltonian $\tH$ is chosen appropriately, a renormalized theory (i.e., a limit of removing the cut-off or, equivalently, a continuum limit) exists and leads to the IBC process introduced here.

Another way of implementing a UV cut-off is to smear out the $x$-particles over (say) a ball of radius $\delta>0$ by means of a (square-integrable) profile function $\varphi: \RRR^3\to\RRR$ that replaces the Dirac delta function (see, e.g., \cite{Der03,DGTZ03,TT15a,ibc2a}). Removing the cut-off corresponds to the limit $\varphi\to\delta^3$. It is known that for a cut-off Hamiltonian $H_\varphi$ analogous to that of Model~2, this limit exists \cite{Der03} and yields the IBC Hamiltonian \cite{ibc2a}. Suppose that $\varphi$ has support of radius $\delta>0$. Since the Bell-type process $(Q^\varphi_t)$ for $H_\varphi$ has an emission rate that vanishes outside the support of $\varphi$, it is clear that it must move the same way as the IBC process $(Q_t)$ when all $y$-particles are outside the $\delta$-ball. Together with the equivariance of $|\psi|^2$ for both processes and the fact that both processes can only jump one sector up or down at a time, this is strong evidence (though not rigorous proof) that $(Q^\varphi_t)$ approaches $(Q_t)$ in the limit.

A third way (not common) of implementing a UV cut-off is to assume that the $x$-particles are spheres of radius $\delta>0$, as described in Remark~\ref{rem:deltasphere} for Model~3. As discussed there, it is plausible to conjecture that the limit of removing the cut-off, $\delta\to0$, exists and yields the IBC Hamiltonian and the IBC process of Model~3.

These examples suggest that renormalization usually leads to Bell-type processes with IBC. It would be of interest to study this question further and obtain a full picture of the conditions under which this is the case.

\subsection{Remarks}
\label{sec:rem}

We conclude this section with a number of remarks.

\begin{enumerate}
\setcounter{enumi}{\theremarks}
\item \textit{Radical topology.} In Remark~\ref{rem:radical} in Section~\ref{sec:M4rem}, we have described the ``radical topology'' for Model~4; an analogous choice of topology is possible for Models~1, 2, and 3. For Model~1, it can be obtained by identifying configurations $(x,y)$ such that $\vy_j=\vx_i$ with $(x,y\setminus \vy_j)$, thus gluing the $(m,n)$-particle sector into the diagonal of the $(m,n+1)$-particle sector. In this topology, the process $(Q_t)$ has continuous paths. This view seems particularly natural when using a different configuration space $\Q$ based on unordered configurations \cite{fermionic} (see Remark~\ref{rem:permutation} below) and ``all particles are identical'' \cite{aapi}. 

\item \textit{Known boundary conditions.} The use of boundary conditions with the Schr\"odinger equation is well known; these are not interior--boundary conditions but concern solely the value and/or derivative of the wave function at every boundary point. The Hamiltonian is a self-adjoint version of the negative Laplace operator whose domain consists of functions satisfying the linear boundary condition. Dirichlet, Neumann, or Robin boundary conditions are often considered on a codimension-1 boundary (such as the end points of an interval) and imply that the probability current vanishes on the boundary, with the consequence that the Bohmian trajectory has probability zero to ever reach the boundary. Also cases in which the boundary has codimension 3 have long been considered, for example in the Bethe--Peierls boundary condition \cite{BP35}
\be\label{BethePeierls}
\lim_{r\searrow 0}\Bigl(\frac{\partial}{\partial r} + \alpha\Bigr) \Bigl(r\psi(r\vomega)\Bigr) =0
\ee
for a wave function $\psi\in L^2(\RRR^3)$. For this condition, which appears in the theory of zero-range interactions (delta potentials), the configuration space is $\RRR^3\setminus\{\vzero\}$, and the boundary is the origin. Also this condition leads to a vanishing current into the boundary, so that the Bohmian trajectory never reaches the origin.

Conditions that allow a non-zero current into the boundary have been considered for graphs (also known as networks), i.e., spaces obtained by gluing several intervals together at their end points. Usually the probability content of a vertex is taken to vanish (which means for the Bohmian particle that it has probability zero to stay at the vertex for a positive duration), which leads to the Kirchhoff condition that at every vertex at every time, the sum of the currents is zero (or, equivalently, the sum of the inward-pointing currents equals that of the outward-pointing ones). One imposes linear vertex conditions on the wave function, which we may call boundary conditions if we think of the vertices as the boundary points of the intervals. The vertex conditions that imply the Kirchhoff condition and lead to a self-adjoint Hamiltonian are known, see, e.g., \cite{KS99,kuchment}; the simplest ones demand that $\psi$ is continuous at every vertex (i.e., the limiting values as the vertex is approached along different edges coincide), and that the sum of the derivatives along each edge at this vertex is proportional, with a given proportionality constant, to the value of $\psi$ at the vertex; so this condition is similar to a Robin boundary condition in that it involves both the value and the derivative of $\psi$. The Bohmian trajectories for such Hamiltonians are discussed in \cite{Tum05} and form a stochastic process similar to Bell-type QFTs: the Bohmian particle moves deterministically according to Bohm's law of motion along an edge until it reaches a vertex; at this instant, the particle chooses randomly the edge along which to leave the vertex, with a probability distribution expressed by a law of the theory in terms of the wave function. A parallel with the IBC processes can be seen in that the arrival at the vertex occurs deterministically (as does the absorption in IBC models), while the departure from the vertex, which of course occurs in the same instant, is stochastic (like the emission in IBC models) in that the direction of departure is random.

\item \textit{Kinks in the trajectories for non-Dirichlet IBCs.} It is a known feature of Bell-type QFTs that, whenever a particle gets created or annihilated, there will generically be a kink in the world line of every particle, corresponding to a discontinuous change in the velocity. That is because, for a jump $q\to q'$ in configuration space corresponding to the creation or annihilation of a particle, there is no reason why the Bohmian velocity $\Im (\nabla \psi/\psi)$ (or any of its components) would have to be the same at $q'$ as at $q$. This phenomenon also occurs for some of the IBC models, such as Model~1 and Model~2 with IBCs of the Neumann or Robin type, see Figure~\ref{fig:kink}. 

\begin{figure}[h]
\begin{center}
\includegraphics[width=0.5\textwidth]{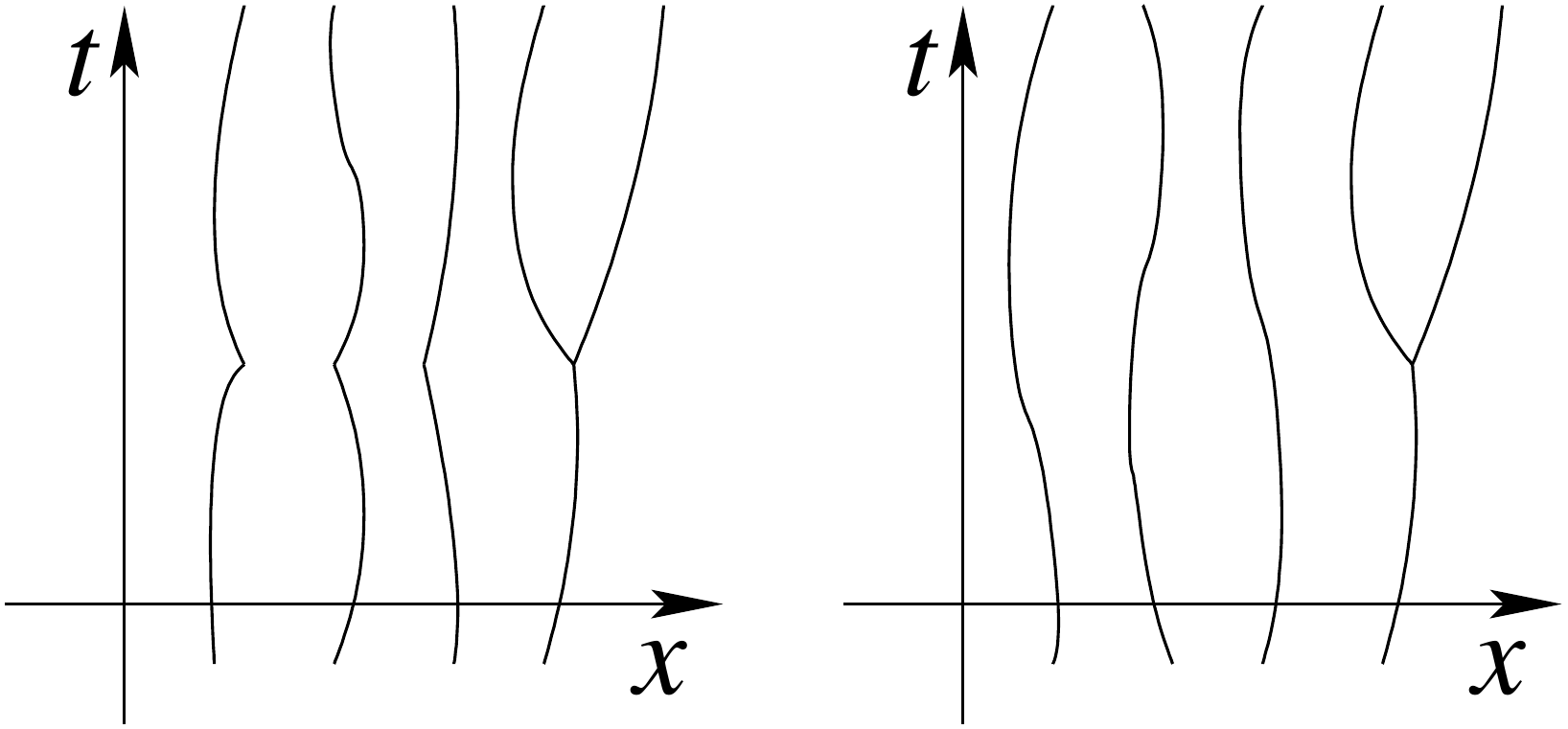}
\end{center}
\caption{In some IBC models, all particles undergo a kink (i.e., a discontinuous change in velocity) at every time of a particle creation or annihilation. LEFT: This is the case for IBCs of the Neumann and Robin type. RIGHT: For IBCs of the Dirichlet type, a kink occurs only for the emitting or absorbing particle.}
\label{fig:kink}
\end{figure}

In contrast, an IBC of the Dirichlet type, such as \eqref{IBC2} or \eqref{IBC1}, implies the absence of such kinks for all particles except the emitting or absorbing one. For the other particles, the velocity does not change at the time of a creation or annihilation, 
but the acceleration jumps, so that the world line ($t\mapsto \vX_i(t)$ or $t\mapsto \vY_j(t)$) is non-smooth (of class $C^1$ but not $C^2$) at the time of a creation or annihilation.
The velocity is continuous because this condition, say \eqref{IBC2} for definiteness, entails that, as the configuration $y\in\Q^{(n)}$ approaches the boundary, say $\vy_j\to\vzero$, the components of $\Im (\nabla\psi^{(n)}/\psi^{(n)})$ parallel to the boundary (i.e., the components for $\vy_k$ with $k\neq j$) approach $\Im (\nabla\psi^{(n-1)}/\psi^{(n-1)})$. 

In Model~2, in which the $x$-particle cannot have a kink because it is fixed at the origin, it follows that none of the particles has a kink at the time of a creation or annihilation; note, however, that the path in configuration space may well have a kink (as shown in Figure~\ref{fig:radical}) when reaching the boundary and the radical topology is used, as the component of the Bohmian velocity \emph{vertical} to the boundary usually jumps from a nonzero value to zero. 

In Model~1, the kink in the world line of the emitting or absorbing $x$-particle arises as follows. Consider, for example, an absorption event. For the velocity of the configuration, the component parallel to the boundary does not jump, and that includes the velocities of all unrelated particles, as well as the center of mass of the absorbing $x$-particle and the absorbed $y$-particle. That is, the velocity of the center of mass just before the absorption equals the velocity of the $x$-particle just after the absorption. Since the relative velocity between the $x$ and the $y$ just before absorption is nonzero, the $x$-trajectory has a kink. So, an $x$-particle undergoes a kind of recoil when it emits or absorbs a $y$-particle.

\item \textit{O'Flanagan's paradox.} This was originally formulated for Model~1 with the kind of UV cut-off described in Remark~\ref{rem:deltasphere}: The configuration consists of those points in $\cup_{m,n} \RRR^{3m}_x \times \RRR^{3n}_y$ for which every $y$-particle has distance at least $\delta>0$ from every $x$-particle. A $y$-particle gets absorbed as soon as the distance reaches $\delta$. O'Flanagan described his puzzle as follows:
\begin{quotation}
``If we think just about the Schr\"odinger evolution
during a process where a photon [$y$-particle] gets absorbed by an electron [$x$-particle],
it seems to me that the information about which direction the
photon came in from is lost. That is, construct an initial
state of the wavefunction which is localised in the bulk of
the (1,1) sector but which is heading towards the boundary
very quickly, so that all of the Bohmian trajectories hit
the boundary (or at least the vast majority). Then the
trajectories jump to the middle of the (1,0) sector where
there is just an electron, so almost all of the $|\psi|^2$ must
end up there too. It seems to me that if we try to reverse this
picture, beginning with a wavefunction localised in the (1,0)
sector, then I can't see how we can get it to ``break the symmetry''
and shoot out the photon in the right direction.'' \cite{oflan}
\end{quotation}
Since the reasoning involves only a single $x$-particle which need not move, and only 0 or 1 $y$-particle, it can be considered just as well in Model~3, in which the configuration space is $\Q=\Q^{(0)}\cup\Q^{(1)}$ with $\Q^{(0)}=\{\emptyset\}$ and $\Q^{(1)}=\RRR^3\setminus B_\delta$ with $B_\delta$ the ball of radius $\delta$ around the origin. For the purpose of the reasoning, consider a large $\delta$, and an initial wave function with $\psi^{(0)}=0$ and $\psi^{(1)}$ a wave packet of size much smaller than $\delta$ moving toward the origin along, say, the $z$ axis. In our words, the reasoning is that since every trajectory jumps to $\Q^{(0)}$ when it hits the $\delta$-sphere, all of the wave packet will be transferred to $\Q^{(0)}$, so that, after a while, the wave function will be entirely in $\Q^{(0)}$; that is, up to an irrelevant global phase factor, $\psi^{(0)}=1$ and $\psi^{(1)}=0$. Now consider the time-reverse history, starting out with the complex-conjugate of the final state, $\psi^{(0)}=1$ and $\psi^{(1)}=0$. By time reversal symmetry, it should evolve into a wave packet (of size much smaller than $\delta$) in $\Q^{(1)}$ moving away from the origin along the $z$ axis. However, the wave function $\psi^{(0)}=1$, $\psi^{(1)}=0$ is rotationally symmetric and contains no information about any direction in space, so it cannot evolve into a packet moving in the $z$ direction. That is the paradox.

The solution is that while it is true that every Bohmian trajectory that reaches $r=\delta$ jumps to $\Q^{(0)}$, it is not the case that the wave packet just disappears into that surface like into the horizon of a black hole. Instead, part of the wave packet actually gets reflected at the surface $r=\delta$ (and some trajectories never reach $r=\delta$). In fact, if we expand the wave packet in spherical harmonics, then only the first term, which is constant over the sphere, couples to $\Q^{(0)}$ via the IBC \eqref{IBC3} and the appropriate term in the Hamiltonian \eqref{H3def1}. So, the wave function does not end up being rotationally symmetric, and in the time-reverse history, the wave function emitted from $\Q^{(0)}$ at $r=\delta$ superposes with the time-reverse of the reflected wave function in $\Q^{(1)}$ to form a small packet moving in the $z$ direction.

If we consider O'Flanagan's paradox for $\delta=0$, then, in addition to reflection, part of the wave packet might evade absorption into $\Q^{(0)}$ by passing around the origin. Thus, the situation seems less paradoxical for $\delta=0$. Rather the other way around, if only the part of the wave function that reaches the origin can be absorbed, then one might worry that nothing will be absorbed, as the origin is a set of measure zero. But we know for fact \cite{ibc2a} that the $n$-particle sector is not invariant under the time evolution for any $n$, so the origin must attract a significant part of the wave function in $\Q^{(1)}$ by ``sucking in'' the wave function immediately adjacent to it.

\item \textit{Location of emission.} When an atom emits a photon, does the photon trajectory start at the electron or at the nucleus? Does the electron jump to a new location, or move there continuously? Is the electron at rest before and after? These questions can be answered in the non-relativistic framework of Model~1 (or rather, a variant thereof), taking photons to be the $y$-particles, electrons the $x$-particles, and introducing $z$-particles for the nuclei, along with a $-1/|\vx_i-\vz_k|$ potential for every $x$ and every $z$. Since both $x$ and $z$ are charged, they should both fundamentally be capable of emitting $y$'s, each with an IBC. Suppose the initial situation comprises a single $x$ and a single $z$, and suppose the initial wave function is an excited eigenstate of the Hamiltonian without $y$-creation (a ``hydrogen eigenstate''); so the $x$ and $z$ are initially at rest relative to each other. Then the wave function created in the sector with one $y$ more can approach the hydrogen ground state, with the excess energy in the $y$. The $x$ and the $z$, if they have equal absolute charge, are equally likely to emit a $y$. None of the particles jumps, they all move continuously. Since the hydrogen ground state has a different position distribution than an excited state, the $x$ and $z$ need to move relative to each other before they settle into their final configuration. The emission of the $y$ probably occurs in the middle of this motion. One further complication can arise from the fact that an $x$ or $z$ is usually surrounded by a cloud of $y$'s; so the newly emitted $y$ may actually settle in the cloud while another $y$ from the cloud may escape to infinity. A more detailed and more careful study of these questions would be of interest. 

\item \textit{Boundaries at infinity.} Some Bohm-like theories, with a different equation of motion than Bohm's, have particles reaching infinity in finite time \cite{SV09}. This problematical behavior can perhaps be dealt with by regarding $|\vx|\to\infty$ as a ``boundary surface'' of $\RRR^3$ (similar to the way infinity appears as a boundary surface in Penrose's conformal diagrams of space-time) and introducing conditions analogous to IBCs to govern the flux into the boundary and terms to ensure that the probability that has flown into the boundary re-appears in a different place.

\item \textit{Question of rigorous existence.} It would be of interest to rigorously prove the existence of the Bell-type process with IBC for all times (``global existence theorem''). Corresponding existence proofs, under suitable conditions, are known for Bohmian mechanics \cite{bmex,TT05} and for the Bell-type process for lattice QFTs \cite{GT05,TG04}; such proofs usually also provide a rigorous proof of equivariance. For IBC processes, such a proof would need to control that there cannot be infinitely many jumps in finite time; and that the trajectory has probability 0 to ever run into a node of the wave function or to escape to infinity.

\item \textit{Question of wave functions outside the domain of $H$.} For defining the process, we have assumed that $\psi$ lies in the domain $\domain$ of the Hamiltonian (initially, and therefore for all times), just like one usually considers Bohmian mechanics for smooth wave functions (which lie in the domain of the corresponding Hamiltonian). However, the unitary time evolution is defined also for wave functions $\psi\in\Hilbert\setminus \domain$, and the question arises whether also for such $\psi$ a process $(Q^\psi_t)$ analogous to the one we have defined here can be defined.
\end{enumerate}
\setcounter{remarks}{\theenumi}

\section{Symmetries}
\label{sec:sym}

In this section, we study for various kinds of symmetries of the Hamiltonian whether the IBC process also respects these symmetries. We find that it does, which provides support for regarding the IBC process as a reasonable picture of physical reality.

\subsection{Several Symmetries}

We cover several kinds of symmetries in the following remarks.

\begin{enumerate}
\setcounter{enumi}{\theremarks}
\item \textit{Time reversal.} 
All of the IBC processes described in this paper are invariant under time reversal, as described in detail for Model~4. We describe in \cite{ST18} how cases can arise in which time-reversal symmetry is violated; as discussed there, these cases involve \emph{complex} coefficients in the IBC [such as $g$ in \eqref{IBC4} and \eqref{H4def} or $e^{i\theta}$ in \eqref{Robin}] that have different phases at different boundaries (or for different $x$-particles).

\item \textit{Global phase factor.}
A basic symmetry of all Bell-type processes with IBC is that if we change the wave function by a global phase factor, $\tilde\psi_t=e^{i\phi} \psi_t$ with $\phi\in\RRR$, then the process associated with $\tilde\psi$ coincides with that associated with $\psi$, as can be seen from the fact that $e^{i\phi}$ cancels out of $|\psi|^2$, out of the Bohmian velocity, and out of the jump rate \eqref{jumprate4}, 
\eqref{jumprate3}, \eqref{jumprate2}, or \eqref{jumprate1}. Since this cancelation would also occur if the phase factor were time-dependent, it follows further that the \textit{addition of a constant $E$ to the Hamiltonian}, which leads to a change in the wave function by a time-dependent global phase factor $e^{-iEt/\hbar}$, has no effect on the process.

\item \textit{Symmetries of Euclidean Space.} 
The IBC process for Model~1 (with moving $x$-particles) is invariant under all symmetries of Euclidean 3-space (translations, rotations, inversion $\vx\mapsto -\vx$, and their composites). This is perhaps rather obvious; it means that if $\psi$ evolves according to $H$ and $(Q_t)_{t\in\RRR}$ is the associated process, then for any $\va\in\RRR^3$ and $R\in O(3)$ (i.e., a rotation including, possibly, a space inversion), $\tilde\psi$ defined by
\be
\tilde\psi_{t}\bigl(\vx_1,\ldots,\vx_m,\vy_1,\ldots,\vy_n\bigr):= \psi_{t}\bigl(R\vx_1+\va, \ldots, R\vx_m+\va, R\vy_1+\va, \ldots,R\vy_n+\va\bigr)
\ee
also evolves according to $H$, and the process $(\tilde Q_t)_{t\in\RRR}$ defined by
\be
\tilde Q_t = T(Q_{t})\,,
\ee
with $T:\Q\to\Q$ defined by
\be
T\bigl(\vX_1,\ldots,\vX_m, \vY_1,\ldots,\vY_n\bigr) = \bigl(R\vX_1+\va, \ldots,R\vX_m+\va, R\vY_1+\va,\ldots, R\vY_n+\va\bigr)\,,
\ee
is the associated process. The model is also invariant under time translations.

\item \textit{Gauge invariance.} 
In Model~1, a gauge transformation does not affect the process $Q_t$. In more detail, suppose we introduce an external magnetic field into Model~1 by replacing all derivatives $\nabla_{\vx_i}$ in \eqref{Bohmx} and \eqref{H1def} by $\nabla_{\vx_i}-i e \vA(\vx_i)$, where $\vA$ is the vector potential and $e$ the electric charge of the $x$-particles. 
Now the gauge transformation corresponding to the function $f:\RRR^3\to\RRR$,
\be
\tilde\vA = \vA +\nabla f
\ee 
and
\begin{equation}
  \tilde\psi(\vx_1,\ldots,\vx_m,\vy_1,\ldots,\vy_n) = e^{i\sum_i e f(\vx_i)} \psi(\vx_1,\ldots,\vx_m,\vy_1,\ldots,\vy_n)\,, 
\end{equation}
leads to a wave function $\tilde\psi$ that obeys the same equations as $\psi$ but with $\tilde\vA$ instead of $\vA$, and the process associated with $\tilde\psi$ coincides with that associated with $\psi$.

\item \textit{Permutation invariance.}\label{rem:permutation} Given that $Q=(\vx_1,\ldots,\vx_m,\vy_1,\ldots,\vy_n)$, any permutation of the $\vx$'s among themselves and the $\vy$'s among themselves will not change the process starting from $Q$ except for the labeling of the particles; this situation is the same as in Bohmian mechanics (without particle creation), discussed in \cite{DGTTZ06,fermionic}. Specifically, let $\tilde{Q}$ be such a permutation of $Q$; the solution of Bohm's equation of motion from initial configuration $\tilde{Q}$ is the same as that from $Q$ up to relabeling, the time at which $\tilde{Q}_t$ hits the boundary is the same as for $Q_t$, the configurations that $\tilde{Q}_t$ and $Q_t$ then jump to are permutations of each other, the rate of emitting a $y$-particle is the same for $\tilde{Q}_t$ as for $Q_t$, and the configurations they jump to in that event are also permutations of each other. Put differently, the Bell-type process with IBC for Model~1 can be defined just as well on the space of \emph{unordered} configurations,
\be
\Bigl\{(q_x,q_y):q_x\subset \RRR^3, q_y \subset\RRR^3, \#q_x<\infty, \#q_y<\infty,q_x\cap q_y=\emptyset\Bigr\}\,.
\ee

\end{enumerate}
\setcounter{remarks}{\theenumi}

\subsection{Galilean Symmetry}

Bohmian mechanics for a fixed number of particles is invariant under Galilean boosts, as discussed in, e.g., \cite{DGZ92,AGTZ08}. However, Model~1, along with the jump process $Q_t$, fails to be Galilean covariant. 
In this section, we elucidate why that is, how Galilean symmetry is related to local conservation of mass, why mass is not conserved in Model~1, and how Model~1 can be so modified as to become invariant under the action of the Galilean group (i.e., the group of symmetries of Galilean space-time; see, e.g., \cite{Arn}). The result, which we will call Model $1'$, is an IBC version of the QFT known as the ``Galilean invariant Lee model'' (GILM) \cite{LL67,Schr68,DD76} (see \cite{Lee54,Schw61} for the original Lee model). As a by-product, our consideration provides an alternative way of arriving at the GILM. We show here that the IBC process that we set up for the GILM is Galilean covariant as well, so the theory satisfies Galileo's principle of relativity. This result shows that the concepts of IBC and Bell-type QFT are compatible with Galilean symmetry. It can also be regarded as providing a deeper justification for the name ``Galilean invariant Lee model.''\footnote{In \cite{DGTZ03}, some of us had written about a variant of Model~1 with UV cut-off that the reason why Galilean boost covariance fails ``is that, roughly speaking, a photon gets created with wave function $\varphi$ which cannot be Galilean invariant. [\ldots] even with the cutoff removed, i.e., for $\varphi(\vy) = \delta(\vy)$, the quantum dynamics [i.e., the evolution of the wave function] is not Galilean invariant.'' These statements are not quite on target because the issue is not so much that the wave function $\varphi$ cannot be Galilean invariant (an issue that would actually disappear when replacing $\varphi(\vy)$ by $\delta(\vy)$ as boosts amount to a time-dependent translation in momentum space and the Fourier transform of $\delta$ is translation invariant) but the non-conservation of mass. In fact, once the mass conservation is restored in the GILM, one can also introduce a UV cut-off without violating Galilean covariance \cite{LL67}.}

In order to show that Model~1 breaks Galilean symmetry, we start from the usual transformation formula for wave functions; it says that 
under a Galilean boost with relative velocity $\vv\in\RRR^3$, i.e., under the space-time coordinate transformation
\be\label{T}
(\tilde t, \tilde \vx) = (t,\vx+\vv t)\,,
\ee
the wave function $\psi:\RRR^{3N}\times \RRR_t \to \CCC$ of $N$ particles in non-relativistic quantum mechanics transforms  according to 
\be\label{Galilean}
  \tilde{\psi}_t(\vx_1, \ldots, \vx_N) = \exp\biggl[\tfrac{i}{\hbar}
  \sum_{i=1}^N m_i
  (\vv\cdot \vx_i  - \tfrac{1}{2} \vv^2 t)\biggr] \, \psi_t(\vx_1-\vv t,
  \ldots, \vx_N-\vv t)\,,
\ee
where $m_i$ is the mass of the $i$-th particle. As we will elucidate, in a setup with particle creation, this formula requires the conservation of mass. However, in Model~1, particles get created and annihilated according to the particle reaction 
\be\label{xxy}
x\rightleftarrows x+y\,.
\ee
Mass is not conserved in this reaction, simply because a particle of mass $m_x$ gets replaced by two particles of total mass $m_x+m_y>m_x$, or vice versa. Mass conservation can be restored by assuming a particle reaction of the form 
\be\label{xzy}
x \leftrightarrows z+y
\ee
with
\be\label{mxmzmy}
m_x=m_z+m_y\,.
\ee
That is, an $x$-particle, when emitting a $y$-particle, mutates into a $z$-particle with lesser mass, while a $z$-particle, when absorbing a $y$-particle, mutates into an $x$-particle with greater mass. Eq.s \eqref{xzy} and \eqref{mxmzmy} summarize the GILM in a nutshell; in the original Lee model \cite{Lee54,Schw61}, $m_x$ and $m_z$ were assumed to be infinite, so that the $x$- and $z$-particles do not move. (If we wish, we can introduce further particle species $w,\ldots$ of mass $m_w=m_z-m_y$ etc.\ (provided that is still positive) and reactions $z \leftrightarrows w+y$ etc.. However, for the sake of simplicity, we will assume here that a $z$-particle cannot emit any particles---it can only absorb a $y$-particle and thereby mutate into an $x$-particle. In particular, $z$'s cannot emit $y$'s, and $x$'s cannot absorb $y$'s.)

\subsubsection{Galilean Covariance and Conservation of Mass}

We now explain in terms of Model~1 why Galilean covariance requires the conservation of mass. (A different reasoning is described in \cite{Bar54}.) The reason is orthogonal to the issue of UV divergence and thus easiest explained in terms of the original, naive, formal Hamiltonian of which Model~1 is a precise interpretation,
\begin{align}
(H_\orig \psi)(x,y) &= -\tfrac{\hbar^2}{2m_x} \sum_{i=1}^m \nabla_{\vx_i}^2 \psi (x,y)
-\tfrac{\hbar^2}{2m_y} \sum_{j=1}^n \nabla_{\vy_j}^2 \psi(x,y)  +nE_0 \psi(x,y) \nonumber\\
&\quad +\: g \sqrt{n+1} \sum_{i=1}^m \psi\bigl(x; (y,\vx_i)\bigr)\nonumber\\
&\quad +\: \frac{g}{\sqrt{n}} \sum_{i=1}^m \sum_{j=1}^n  \delta^3(\vx_i-\vy_j)\,\psi\bigl(x,y\setminus \vy_j\bigr)\,.\label{Horigdef1}
\end{align}
As an abbreviation, let 
\be
T_t(\vx)= \vx+\vv t\,, ~~~~
T_t(x,y) = \bigl(T_t\vx_1,\ldots,T_t\vx_m,T_t \vy_1,\ldots,T_t\vy_n\bigr)\,.
\ee
The obvious extension of the transformation law \eqref{Galilean} to a wave function $\psi$ on the configuration space $\Q= \cup_{m,n} \RRR_x^{3m} \times \RRR_y^{3n}$ of a variable number of particles reads
\be
  \tilde{\psi}_t(x,y) = e^{i\alpha_t(x,y)}
   \, \psi_t\bigl(T_t^{-1}(x,y)\bigr)
\ee
with the additional phase $\alpha$ given by
\be
\alpha_t(x,y) = \frac{1}{\hbar}\vv\cdot \Bigl(\sum_{i=1}^m m_x\vx_i + \sum_{j=1}^n m_y \vy_j\Bigr)
  -\frac{M}{2\hbar} \vv^2 t\,,
\ee
where
\be
M= mm_x + nm_y
\ee
is the total mass of all particles in the configuration $(x,y)$.
The fact that the difference
\be
\alpha_t(x,(y,\vx_i)) - \alpha_t(x,y) = \frac{1}{\hbar} \vv \cdot m_y \vx_i - \frac{m_y}{2\hbar} \vv^2 t 
\ee
is nonzero has the consequence that the second line in \eqref{Horigdef1} transforms differently than it should: It implies that the Schr\"odinger equation
\be\label{Schr}
i\hbar\partial_t\psi = H\psi
\ee
does not hold for $\tilde\psi$ if $\vv\neq 0$. A problem of the same type arises from the third line of \eqref{Horigdef1}. This is the point where Galilean covariance fails.

In order to obtain Galilean covariance, we need that the additional phase $\alpha$ is the same at any two configurations related by a creation or annihilation event. This is equivalent to saying that the mass is conserved during particle creation and annihilation.

\subsubsection{Galilean Covariant Model}
\label{sec:1prime}

We thus consider a variant of Model~1, which we will call Model $1'$ and in which the configuration space is
\be\label{confdef}
\Q = \bigcup_{m=0}^\infty \bigcup_{n=0}^\infty \bigcup_{\ell=0}^\infty \Q^{(m,n,\ell)} 
= \bigcup_{m=0}^\infty \bigcup_{n=0}^\infty \bigcup_{\ell=0}^\infty (\RRR^3_{\vx})^m \times (\RRR^3_{\vy})^n \times (\RRR^3_{\vz})^\ell\,,
\ee
so that the wave function $\psi_t:\Q\to\CCC$ is of the form $\psi(x,y,z)$ with $x=(\vx_1,\ldots,\vx_m)$, $y=(\vy_1,\ldots,\vy_n)$, and $z=(\vz_1,\ldots,\vz_\ell)$; $\psi$ is symmetric against permutation of $\vy$'s, anti-symmetric against permutation of $\vx$'s, and anti-symmetric against permutation of $\vz$'s. The formal Hamiltonian of Model $1'$ is given by
\begin{align}
(H_\orig \psi)(x,y,z) &= -\tfrac{\hbar^2}{2m_x} \sum_{i=1}^m \nabla_{\vx_i}^2 \psi
-\tfrac{\hbar^2}{2m_y} \sum_{j=1}^n \nabla_{\vy_j}^2 \psi
-\tfrac{\hbar^2}{2m_z} \sum_{k=1}^\ell \nabla_{\vz_k}^2 \psi 
+nE_0 \psi \nonumber\\
& +\:  \tfrac{g\sqrt{(n+1)(\ell+1)}}{\sqrt{m}}\sum_{i=1}^{m} (-1)^{\ell+i+1} \psi\bigl(x\setminus \vx_i; (y,\vx_i);(z,\vx_i)\bigr)\nonumber\\
& +\: \tfrac{g\sqrt{m+1}}{\sqrt{n\ell}} \sum_{j=1}^n  \sum_{k=1}^\ell (-1)^{m+k+1} \delta^3(\vz_k-\vy_j)\,\psi\bigl((x,\vz_k);y\setminus \vy_j;z\setminus \vz_k\bigr)\,.\label{H1primeorigdef}
\end{align}
The transformation law reads
\be\label{tildepsi}
  \tilde{\psi}_t(x,y,z) = e^{i\alpha_t(x,y,z)}
   \, \psi_t\bigl(T_t^{-1}(x,y,z)\bigr)
\ee
with 
\be
T_t(x,y,z) = \bigl(T_t\vx_1,\ldots,T_t\vx_m,T_t \vy_1,\ldots,T_t\vy_n,T_t\vz_1,\ldots,T_t\vz_\ell\bigr)\,.
\ee
and the additional phase $\alpha$ given by
\be
\alpha_t(x,y,z) = \frac{M}{\hbar}\vv\cdot \vq
  -\frac{M}{2\hbar} \vv^2 t\,,
\ee
where $\vq$ is the center of mass of the configuration $(x,y,z)$,
\be
\vq =  \frac{1}{M}\Bigl(\sum_{i=1}^m m_x\vx_i + \sum_{j=1}^n m_y \vy_j + \sum_{k=1}^\ell m_z \vz_k\Bigr)\,,
\ee
and $M$ is the total mass of all particles in this configuration,
\be
M= mm_x + nm_y + \ell m_z\,.
\ee
Note that any configuration obtained from $(x,y,z)$ through an emission or absorption event, i.e., $\bigl(x\setminus \vx_i; (y,\vx_i);(z,\vx_i) \bigr)$ or $\bigl((x,\vz_k);y\setminus \vy_j;z\setminus \vz_k\bigr)$ when $\vy_j=\vz_k$, has the same center of mass $\vq$ and the same total mass $M$, and thus the same $\alpha_t$, as $(x,y,z)$. Also, permutations (of the $x$-particles among themselves etc.)\ do not change $\alpha_t$.

Now a straightforward calculation shows that the Schr\"odinger equation \eqref{Schr} for $\psi$ implies that for $\tilde\psi$ on the non-rigorous level.

\subsubsection{IBC and Corresponding Hamiltonian}
\label{sec:GalileanIBC}

We now describe an IBC version of Model $1'$, which is believed to also rigorously define a Hamiltonian \cite{Lam18,Lam18b}. The IBC is given by
\be\label{IBC1prime}
\lim_{(\vy_j,\vz_k)\to(\vx,\vx)} \,|\vy_j-\vz_k| \,\psi(x,y,z)
= \frac{g\mu_{yz}}{2\pi\hbar^2\sqrt{n}}\, \psi\bigl((x,\vx);(y\setminus \vy_j);(z\setminus \vz_k)\bigr)\,,
\ee
where $\mu_{yz}$ is the harmonic mean of $m_y$ and $m_z$,
\be
\mu_{yz} = \frac{m_ym_z}{m_z+m_y}\,.
\ee
The corresponding Hamiltonian is given by
\begin{align}
&(H\psi)(x,y,z) 
= -\tfrac{\hbar^2}{2m_x} \sum_{i=1}^{m} \nabla^2_{\vx_i}\psi
-\tfrac{\hbar^2}{2m_y} \sum_{j=1}^{n} \nabla^2_{\vy_j}\psi-\tfrac{\hbar^2}{2m_z} \sum_{k=1}^{\ell} \nabla^2_{\vz_k}\psi + nE_0 \psi\nonumber\\
& ~~~~~+ \tfrac{g\sqrt{(n+1)(\ell+1)}}{\sqrt{m}}\sum_{i=1}^{m} (-1)^{\ell+i+1} c_{0,i}(x,y,z)
\nonumber\\
& ~~~~~+\: \tfrac{g\sqrt{m+1}}{\sqrt{n\ell}} \sum_{j=1}^n  \sum_{k=1}^\ell (-1)^{m+k+1} \delta^3(\vz_k-\vy_j)\,\psi\bigl(x,\vz_k;y\setminus \vy_j;z\setminus \vz_k\bigr)\label{H1primedef}
\end{align}

The transformation law is still given by \eqref{tildepsi}. Since two configurations connected by the reaction $x \leftrightarrows z+y$, viz., $(x,y,z)$ and $\bigl(x\setminus \vx_i;(y,\vx_i);(z,\vx_i)\bigr)$, have the same $\vq$ and $M$ and thus the same $\alpha_t$, $\tilde\psi_t$ satisfies the IBC \eqref{IBC1prime} if $\psi_t$ does and evolves according to $H$ if $\psi_t$ does.

\subsubsection{Jump Process}
\label{sec:Galileanjump}

For Model $1'$, this process is defined analogously to that for Model~1 in Section~\ref{sec:M1}. Between jumps, the configuration $Q_t$ moves according to Bohm's equation of motion. 
Whenever a $y$-particle and a $z$-particle meet, they get replaced by an $x$-particle. Conversely, the $i$-th $x$-particle spontaneously decays into a $y$ and $z$ (departing in directions $\vomega$ and $-\vomega$, respectively) with rate
\be\label{jumprate1prime}
\sigma_t \;  d^2 \vomega 
= \tfrac{\hbar}{\mu_{yz}} \frac{\Im^+ [c_{-1,i}^*(x,y,z) \, c_{0,i}(x,y,z)]}{|\psi(x,y,z)|^2} d^2\vomega\,.
\ee
This completes the definition of the process $(Q_t)_{t\geq t_0}$ for arbitrary initial time $t_0$. Since for two different choices of $t_0$, $t'_0<t''_0$, the process $(Q''_t)_{t\geq t''_0}$ has the same distribution as the restriction of the process $(Q'_t)_{t\geq t'_0}$ to times $t\geq t''_0$, all the processes obtained for any choice of $t_0$ fit together to form a process $(Q_t)_{t\in\RRR}$ for all times.

We now state the covariance under Galilean boosts:
{\it The process $(\tilde Q_t)_{t\in\RRR}$ defined by
\be
\tilde Q_t = T_t(Q_t)
\ee
has the same distribution as the process associated with $\tilde\psi$.} Let us say this in more detail. A Galilean boost is a mapping of Galilean space-time to itself and therefore maps any particle world line to another world line, and any path $\RRR\to\Q$ in configuration space (which corresponds to several world lines in space-time) to another path in configuration space; it therefore maps any probability distribution over the set of paths in $\Q$ (i.e., any stochastic process in $\Q$) to another such distribution (i.e., another stochastic process). The claim is that the process $(Q_t)_{t\in\RRR}$ associated with the wave function $\psi$ gets mapped to the process associated with a suitable wave function $\tilde\psi$; in fact, $\tilde\psi$ is just the transformed wave function given by \eqref{tildepsi}.

Indeed, this follows from the following four facts: (i)~That $\tilde\psi$ obeys the Schr\"odinger equation with $H$ given by \eqref{H1primedef}, as discussed in Section~\ref{sec:GalileanIBC}. (ii)~The well known fact that Bohm's equation of motion transforms in the right way. (iii)~The jumps $z+y\to x$ do not cause a problem. (iv)~The jump rate \eqref{jumprate1prime} for the reaction $x\to z+y$ is the same for $\tilde\psi_t$ at $T_t(x,y,z)$ as for $\psi_t$ at $(x,y,z)$. 
This is because, when passing from $\psi$ to $\tilde\psi$, both $c_{-1,i}$ and $c_{0,i}$ change by the same phase factor $\exp(i\alpha_t(x,y,z))$.

We have already discussed invariance under space and time translations, reflections, and time reversal, so that the time evolution is invariant under the full Galilean group.

In a sense, the Galilean covariance of the Bohmian version of the theory provides the full justification of the claim of Galilean covariance of the IBC Hamiltonian. For trajectories in space-time (and thus also those in the configuration space considered here), it is unambiguous how they transform under an element $g$ of the Galilean group; it is not just any old action of the Galilean group, it is \emph{the} action that \emph{must} be applied if the paths are to be taken seriously as the paths of particles in space. In contrast, the transformation behavior of the wave function is not per se prescribed; the ultimate reason why \eqref{tildepsi} is the correct transformation behavior under boosts is that the jump process associated with $\tilde\psi$ is the transform of the jump process associated with $\psi$.

\section{Comparison with Known Bell-Type QFTs}
\label{sec:Bell-type}

The IBC processes $Q_t$ defined in previous sections have striking similarities with the process provided by Bell-type QFTs with UV cut-off, which is why we call them Bell-type processes with IBC. The similarities include that both processes have jumps; that the pieces between the jumps are solutions of Bohm's equation of motion; that the configuration is $|\psi|^2$ distributed at every time; that both theories are time reversible; and that the jump rates are governed by a law involving the wave function $\psi$. More specifically, since the jump rate in a Bell-type QFT with UV cut-off is given by \cite{DGTZ03,DGTZ04,DGTZ05a,DGTZ05b}
\be\label{jumprate0a}
\sigma_t(q'\to dq) = \tfrac{2}{\hbar}\frac{\Im^+ \scp{\psi}{P(dq)H_IP(dq')|\psi}}{\scp{\psi}{P(dq')|\psi}}
\ee
with $P(\cdot)$ the configuration operators (a projection-valued measure (PVM) on configuration space acting on Hilbert space) and $H_I$ the interaction Hamiltonian,
or, in the notation of the present paper, by
\be\label{jumprate0}
\sigma_t(q'\to dq) = \tfrac{2}{\hbar}\frac{\Im^+ \bigl[ \psi(q)^* \, \scp{q}{H_I|q'}\, \psi(q') \bigr]}{|\psi(q')|^2}\, dq\,,
\ee
it is striking that both jump rate formulas, \eqref{jumprate0} and, say, \eqref{jumprate4} for Model~4, are of the form
\be
\sigma_t(q'\to dq) = \frac{\Im^+ A(q,q')}{|\psi(q')|^2}\, dq\,,
\ee
with $A(q,q')$ a complex-valued sesqui-linear expression in $\psi$ that is linear in $\psi(q')$ and conjugate-linear in $\psi(q)$; viz., for \eqref{jumprate0},
\be
A(q,q') = \tfrac{2}{\hbar} \, \psi(q)^* \, \scp{q}{H_I|q'} \, \psi(q') \,,
\ee
and for \eqref{jumprate4}, using the IBC \eqref{IBC4},
\be
A(q,q') = \begin{cases}
-\tfrac{2g}{\hbar}\, \delta^2(q-(x,0))\, (\partial_y \psi(q)^*) \, \psi(q') & \text{if }q'=x\in \Q^{(1)}\\
\infty \, \delta(q-x) & \text{if }q'=(x,0)\in\Q^{(2)}\,. 
\end{cases}
\ee
In contrast to \eqref{jumprate4}, the jump rate \eqref{jumprate0} is not uniquely determined by reversibility, equivariance of $|\psi|^2$, and the Markov property. Rather, it is uniquely selected by demanding in addition (i)~the \emph{standard current property} \cite{DGTZ05a,DGTZ05b} that  the probability currents associated with the jumps agree with the standard quantum-mechanical formula, and (ii)~the minimality property that jump rates are as low as possible.

It would be of interest to have a general definition or construction of Bell-type processes that includes the old ones for QFTs with cut-off as well as the IBC processes. 
Another connection between the two is that the Bell-type process with IBC can be regarded as a limiting case of a Bell-type process on a lattice; this is what we turn to now.

\section{Process on a Lattice}
\label{sec:lattice}

Bell \cite{Bell86} originally proposed his jump process for QFTs on a lattice. We now outline how Model~4 can be discretized, why the corresponding process is a special case of Bell's jump process, and how a suitable continuum limit can be taken that leads back to the IBC process.
It is well known \cite{Sudbery,Vink,Col03a,Vi17} (though not on a rigorous
level) that Bohmian mechanics is a limiting case of Bell's lattice process.

A natural discretization of Model~4 goes as follows. Choose a lattice width $\varepsilon>0$. The discretized configuration space $\tQ$ is the union of $\tQ^{(1)}=\varepsilon \ZZZ$ and $\tQ^{(2)}=\{(x,y)\in \varepsilon\ZZZ^2: y> 0\}$, a countable set, with measure $\tmu$ given by $\tmu(\{x\})=\varepsilon$ and $\tmu(\{(x,y)\})=\varepsilon^2$. The Hilbert space $\Hilbert=L^2(\tQ,\CCC,\tmu)$ has inner product
\be
\scp{\psi}{\phi} = \sum_{x\in\tQ^{(1)}} \varepsilon\, \psi^{(1)}(x)^* \, \phi^{(1)}(x) + \sum_{(x,y)\in \tQ^{(2)}} \varepsilon^2\, \psi^{(2)}(x,y)^* \, \phi^{(2)}(x,y)\,.
\ee
We obtain the discrete Hamiltonian $\tH$ from \eqref{H4def} by means of several replacements: replace the second derivative $\partial^2$ by the lattice Laplacian
\be\label{discreteLaplace}
\tLaplace \psi = \varepsilon^{-2} \Bigl( \psi(x+\varepsilon)-2\psi(x)+\psi(x-\varepsilon) \Bigr)\,,
\ee
replace the first derivative $\partial_y \psi^{(2)}(x,0)$ by
\be
\varepsilon^{-1} \Bigl( \psi^{(2)}(x,\varepsilon) - \psi^{(2)}(x,0)\Bigr)\,,
\ee
and replace $\psi^{(2)}(x,0)$ by $(2mg/\hbar^2) \psi^{(1)}(x)$, thus building in the IBC \eqref{IBC4}. That is,
\begin{subequations}\label{tHdef}
\begin{align}
(\tH\psi)^{(1)}(x) 
&= -\tfrac{\hbar^2}{2m}\tLaplace_x \psi^{(1)}
- \tfrac{g}{\varepsilon} \psi^{(2)}(x,\varepsilon) + \tfrac{2mg^2}{\hbar^2\varepsilon} \psi^{(1)}(x) \label{tHdef1}\\[3mm]
(\tH\psi)^{(2)}(x,y) 
&= -\tfrac{\hbar^2}{2m}(\tLaplace_{x}+\tLaplace_{y})\psi^{(2)} \quad \text{for }y>\varepsilon\label{tHdef2}\\[2mm]
(\tH\psi)^{(2)}(x,\varepsilon)
&=-\tfrac{\hbar^2}{2m} \Bigl( \tLaplace_x\psi^{(2)}(x,\varepsilon) +\tfrac{1}{\varepsilon^2} \psi^{(2)}(x,2\varepsilon) - \tfrac{2}{\varepsilon^2}\psi^{(2)}(x,\varepsilon) + \tfrac{2mg}{\hbar^2\varepsilon^2}\psi^{(1)}(x)  \Bigr)\,.\label{tHdef3}
\end{align}
\end{subequations}
The Hamiltonian $\tH$ is formally self-adjoint, and the corresponding Bell process $(\widetilde{Q}_t)$ jumps at random times from one lattice site to another. Given $\widetilde{Q}_t=q'\in\tQ$, the rate of jumping to $q\in\tQ$ is, in agreement with \eqref{jumprate0a} for $H_I=\tH$, given by
\be\label{jumprate0b}
\sigma_t(q'\to q) = \tfrac{2}{\hbar}\frac{\Im^+ \bigl[ \psi(q)^* \, \scp{q}{\tH|q'} \, \psi(q') \bigr]}{|\psi(q')|^2 \,\tmu(\{q'\})}
\ee
with $|q'\rangle$ the Kronecker delta function at $q'$.

Now consider the continuum limit $\varepsilon\to 0$. In $\tQ^{(2)}$ away from the boundary, the process converges to a solution of Bohm's equation of motion. When $\widetilde{Q}_t$ reaches $(x,\varepsilon)\in\tQ^{(2)}$, the rate of jumping to $x\in\tQ^{(1)}$ is
\be\label{jumpratetox}
\sigma_t\bigl((x,\varepsilon)\to x\bigr) = \tfrac{2}{\hbar}\frac{\Im^+ \bigl[ \psi^{(1)}(x)^* \, (-g) \, \psi^{(2)}(x,\varepsilon) \bigr]}{|\psi^{(2)}(x,\varepsilon)|^2 \, \varepsilon^2}\,.
\ee
Since $\psi^{(2)}(x,\varepsilon)-\tfrac{2mg}{\hbar^2}\psi^{(1)}(x)$ approximates $\varepsilon \partial_y\psi^{(2)}(x,0)$ and hence is of order $\varepsilon$, the numerator of the right-hand side of \eqref{jumpratetox} is of order $\varepsilon$, so \eqref{jumpratetox} is of order $\varepsilon^{-1}$, so the waiting time is of order $\varepsilon$, so in the limit $\widetilde{Q}_t$ immediately jumps to $\Q^{(1)}$ when reaching the boundary of $\Q^{(2)}$. Once $\widetilde{Q}_t$ is in $\tQ^{(1)}$, it can jump either along $\tQ^{(1)}$ or to $\tQ^{(2)}$. The matrix elements $\scp{q}{\tH|x}$ with $q\in\Q^{(1)}$ come from the $\tLaplace_x$ in \eqref{tHdef1}, are nonzero only for nearest neighbors, and lead to Bohm's equation of motion in the limit, as long as no jump to $\tQ^{(2)}$ occurs. Given $\widetilde{Q}_t=x\in\tQ^{(1)}$, the only possible jump to $\tQ^{(2)}$ is one to $(x,\varepsilon)$, and the rate for that is
\be
\sigma_t\bigl(x\to(x,\varepsilon)\bigr) = \tfrac{2}{\hbar}\frac{\Im^+ \bigl[ \psi^{(2)}(x,\varepsilon)^* \, (-g) \, \psi^{(1)}(x) \bigr]}{|\psi^{(1)}(x)|^2\, \varepsilon}\,,
\ee
where the numerator is again of order $\varepsilon$, so the jump rate is of order 1, with the consequence that the jump remains stochastic in the limit. To compute the limiting rate, note that
\begin{align}
\psi^{(2)}(x,\varepsilon) 
&\approx \psi^{(2)}(x,0)+\varepsilon \, \partial_y\psi^{(2)}(x,0)\\
&= \tfrac{2mg}{\hbar^2} \psi^{(1)}(x) +\varepsilon \, \partial_y\psi^{(2)}(x,0)\,,
\end{align}
so
\be
\lim_{\varepsilon\to0} \sigma_t\bigl(x\to(x,\varepsilon)\bigr)
= \tfrac{2}{\hbar}\frac{\Im^+ \bigl[ \partial_y\psi^{(2)}(x,0)^* \, (-g) \, \psi^{(1)}(x) \bigr]}{|\psi^{(1)}(x)|^2}\,,
\ee
in agreement with the jump rate \eqref{jumprate4} of the IBC process. This completes our reasoning to the effect that the continuum limit of Bell's lattice process for the Hamiltonian $\tH$ is the IBC process of Section~\ref{sec:Q4def}.

\section{General Setup for Codimension-1 Boundaries}
\label{sec:co1}

In previous sections, we defined the IBC processes for several specific models. Now, in contrast, we give a general definition of the IBC process, if only for the case in which the boundary has codimension 1; a preliminary description was provided in \cite{TG04}.

\subsection{For Schr\"odinger Operators}

We use the setup and notation described in \cite{IBCco1}, where the general IBC for a given configuration space with boundary of codimension 1 was formulated, taking the configuration space $\Q$ to be a finite or countable union of disjoint Riemannian manifolds with boundary (with the masses included in the Riemannian metric), and the wave functions to be cross-sections of some Hermitian vector bundle $E$ over $\Q$. 

We now specify the process for the IBC and Hamiltonian given in \cite{IBCco1}. The process will again be such that the configuration point, when hitting a boundary at $q'\in\partial \Q$, jumps to an interior point $q=f(q')$ in a different sector.  
The Bohmian equation of motion reads:
\be\label{Bohmco1}
\frac{dQ}{dt} = \hbar \,  \frac{\Im\,\bigl( \psi(q), \nabla\psi(q) \bigr)_q}{\bigl(\psi(q),\psi(q)\bigr)_q} (q=Q_t)\,.
\ee
(Here the gradient $\nabla\psi$ is the $E$-valued vector field obtained from the $E$-valued 1-form that is the covariant derivative of $\psi$ by ``raising the index'' using the Riemann metric \cite{DGTTZ06,DGTTZ07}, and $(\cdot,\cdot)_q$ is the inner product in $E_q$.) 
The jump rate formula reads: If $f(q')=q$, then
\be\label{jumprateco1}
\sigma_t(q\to dq') = \hbar \, \frac{\Im^+ \bigl(\psi(q'), \partial_n \psi(q') \bigr)_{q'}}{\bigl(\psi(q),\psi(q)\bigr)_{q}} \nu_{q}(dq')\,,
\ee
where $\nu_{q}$ is the measure on $f^{-1}(q)$ defined by the Riemannian metric \cite{IBCco1}.
This completes the definition of the IBC process $(Q_t)$. The Hamiltonian is formally self-adjoint, and the $|\psi|^2$ distribution is equivariant, as can be seen by a calculation analogous to that of Section~\ref{sec:equiv} (using the known corresponding result \cite{DGTTZ06,DGTTZ07} for Bohmian mechanics in a Riemannian manifold without boundary).

\bigskip

\noindent{\bf Remark.}

\begin{enumerate}
\setcounter{enumi}{\theremarks}
\item \textit{Heisenberg picture vs.\ Schr\"odinger picture.} 
We have formulated our models in the Schr\"odinger picture, regarding vectors in Hilbert space as functions $\psi$ on configuration space $\Q$ that change over time. This is perhaps the simplest and most natural view, but not the only possible. We can also define the process $Q_t$ in terms of a vector $|\psi\rangle$ in an abstract Hilbert space $\Hilbert$, a Hamiltonian $H$ that is a self-adjoint operator on $\Hilbert$, and a projection-valued measure (PVM) $P$ on $\Q$ acting on $\Hilbert$ playing the role of the position (or configuration) operators; in case $\Hilbert=L^2(\Q)$, the configuration PVM is provided by $P(dq)=|q\rangle\langle q| \, dq$. 
In terms of $\Hilbert$, $|\psi\rangle$, $H$, and $P$,
Bohm's equation of motion can be reformulated \cite{DGTZ05b} as follows: For a coordinate function $z:\Q\to\RRR$,
\begin{equation} \label{BohmHeisenberg}
  \frac{dz(Q_t)}{dt} = - \tfrac{1}{\hbar} \frac{\Im \scp{\psi}{P(dq)
  [H,\int_{\Q} z(q') P(dq')]|\psi}}{\scp{\psi}{P(dq)|\psi}} (q=Q(t))\,.
\end{equation}
Likewise, the jump rate law \eqref{jumprateco1} can be written, for $q\in\partial\Q$, as
\be\label{jumprateHeisenberg}
\sigma_t(f(q)\to q) = - \tfrac{1}{\hbar} \, \frac{\Im^+ \scp{\psi}{P(dq) [H,\int_{\Q}\mathrm{dist}(q'',\partial \Q)\, P(dq'')]| \psi}}{\scp{\psi}{P(f(dq))|\psi}}
\ee
with the function $\mathrm{dist}(q'',\partial \Q)$ = distance of $q''$ from $\partial \Q$. The formulas \eqref{BohmHeisenberg} and \eqref{jumprateHeisenberg} are neutral towards the Schr\"odinger versus Heisenberg picture, and can be applied as well in the Heisenberg picture.
\end{enumerate}
\setcounter{remarks}{\theenumi}

\subsection{For Dirac Operators}
\label{sec:Dirac} 

When the Dirac equation is used instead of the non-relativistic Schr\"odinger equation, that is, the Dirac Hamiltonian $H_1=-ic\hbar \valpha\cdot \nabla + \beta m c^2$ instead of the Laplace operator $-(\hbar^2/2m)\nabla^2$, then IBCs can be set up as well \cite{IBCdiracCo1}, and a Bohm-type process $Q_t$ can be defined in much the same way as in the previous sections, as we elucidate now for the case of codimension-1 boundaries. The general definition of the jumps is still as follows: Whenever the Bohmian trajectory in configuration space $\Q$, defined by \cite{Bohm53}
\be
\frac{dQ}{dt} = \frac{j(q)}{\bigl( \psi(q),\psi(q) \bigr)_q}(q=Q_t)
\ee
with $j$ the probability current, reaches the boundary $\partial\Q$ at some point $q'\in\partial \Q$, it jumps to $f(q')$. Conversely, from any interior point $Q_t=q$, the process spontaneously jumps to a boundary point $q'\in f^{-1}(q)$ with rate
\be
\sigma_t(q \to dq') = \frac{j_n(q')^+}{\bigl( \psi(q),\psi(q) \bigr)_q} \nu_q(dq')\,,
\ee
in analogy to the formula \eqref{jumprateco1} for the non-relativistic Schr\"odinger case. Here, $j_n$ means the component inward-normal to the boundary of the probability current, and $\nu_q$ the measure on $f^{-1}(q)$ defined by the Riemannian metric.
It follows in the same way as in the non-relativistic case (see Sections \ref{sec:equiv} and \ref{sec:co1}) that $|\psi|^2$ is equivariant.
Explicit formulas for $H$, $j$, and the IBC are given in \cite{IBCdiracCo1}.

\bigskip

\noindent\textit{Acknowledgments.} 
We thank Stefan Keppeler, Jonas Lampart, Ruadhan O'Flanagan, and Julian Schmidt for helpful discussions.

\end{document}